\documentclass[aps, prd, 10pt, notitlepage, superscriptaddress, nofootinbib,numbers,showpacs,twocolumn]{revtex4-1}

\usepackage[utf8]{inputenc}
\usepackage[T1]{fontenc}
\usepackage{anyfontsize}
\usepackage{amsmath}
\usepackage{amssymb}
\usepackage{amsfonts}
\usepackage{mathrsfs}
\usepackage{bm} 
\usepackage{microtype}

\usepackage{graphicx}
\usepackage{epsfig}

\usepackage[dvipsnames]{xcolor}
\usepackage{hyperref}
\hypersetup{
    colorlinks=true,
    citecolor=Purple,
    linkcolor=Purple,
    urlcolor=Purple,
    linktocpage=true,
    breaklinks=true
}

\usepackage{orcidlink}

\usepackage[capitalize]{cleveref}
\begin{document}
\title{
Black bounce as a quantum correction from string T-duality: Thermodynamics, energy conditions, and observational imprints from EHT}
\author{G. Alencar}
\email{geova@fisica.ufc.br}
\affiliation{Departamento de F\'isica, Universidade Federal do Cear\'a, Caixa Postal 6030, Campus do Pici, 60455-760 Fortaleza, Cear\'a, Brazil.}

\author{T. M. Crispim}
	\email{tiago.crispim@fisica.ufc.br}
	\affiliation{Departamento de F\'isica, Universidade Federal do Cear\'a, Caixa Postal 6030, Campus do Pici, 60455-760 Fortaleza, Cear\'a, Brazil.}

\author{Diego S\'aez-Chill\'on G\'omez}
\email{diego.saez@uva.es} 
\affiliation{Department of Theoretical Physics, Atomic and Optics, and Laboratory for Disruptive Interdisciplinary Science (LaDIS), Campus Miguel Delibes, \\ University of Valladolid UVA, Paseo Bel\'en, 7,
47011 - Valladolid, Spain}
\affiliation{Departamento de F\'isica, Universidade Federal do Cear\'a, Caixa Postal 6030, Campus do Pici, 60455-760 Fortaleza, Cear\'a, Brazil.}
\author{Marcos V. de S. Silva}
	\email{marcosvinicius@fisica.ufc.br}
	\affiliation{Departamento de F\'isica, Universidade Federal do Cear\'a, Caixa Postal 6030, Campus do Pici, 60455-760 Fortaleza, Cear\'a, Brazil.}
    \affiliation{Department of Theoretical Physics, Atomic and Optics, and Laboratory for Disruptive Interdisciplinary Science (LaDIS), Campus Miguel Delibes, \\ University of Valladolid UVA, Paseo Bel\'en, 7,
47011 - Valladolid, Spain}
\date{today; \LaTeX-ed \today}
\begin{abstract}
Motivated by quantum gravity effects suggested by string theory, we investigate gravitational configurations sourced by an effective energy density inspired by T-duality. This density naturally introduces a minimal length scale $l_0$ that acts as an ultraviolet regulator, allowing the description of nonsingular geometries within a classical framework. By employing it as the matter source in the Einstein equations, we construct static and spherically symmetric spacetimes that interpolate smoothly between regular black holes and traversable wormholes, providing a geometric realization of the black bounce scenario. We examine the curvature invariants and confirm the absence of curvature singularities throughout the spacetime. The conditions for the existence of event horizons are analyzed in detail, which allows us to determine the causal structure of the solution. A comprehensive study of the geodesic motion is performed for both massive and massless particles, revealing the presence of photon circular orbits and an innermost stable circular orbit for massive particles. Using observational data from the Event Horizon Telescope, we constrain the minimal length parameter through the black hole shadow radius, finding that for $l_0 \lesssim 1.15\, M_{\text{ADM}}$ our solution remains consistent with observations within the $2\sigma$ confidence level. The optical appearance of spacetime is further investigated by considering a thin accretion disk surrounding the black bounce. From the heat capacity, we analyze the thermodynamic stability of the solution and identify the presence of a phase transition. Finally, we examine the energy conditions and discuss which of them are violated by the effective fluid supporting this geometry.
\end{abstract}
\maketitle
\def\HMS{{\scriptscriptstyle{HMS}}}
\section{Introduction}
\label{S:intro}

Since the first detection of gravitational waves and the advent of other high-precision measurements, the scientific community's attention has increasingly focused on addressing long-standing problems in general relativity, such as the nature of dark energy, the origin of dark matter, and the singularity problem in black holes. These challenges not only motivate the development of new observational techniques but also demand the refinement and extension of our current theoretical frameworks\cite{LIGOScientific:2016aoc, EventHorizonTelescope:2022wkp, EventHorizonTelescope:2019dse, LIGOScientific:2017vwq,Planck:2018vyg,ACT:2020gnv,SDSS:2006srq, DeFalco2020, DeFalco2023}.

It is believed that a quantum gravity theory will solve some of the above problems. One of the key ingredients in such theories is the emergence of a minimum length. This implies that processes with large energies must be suppressed. From the particle path integral viewpoint, large variations of the action cancel, and, to suppress the contributions of lengths below $l_0$, we should expect a contribution $\propto 1/ds$. To implement this, Padmanabhan, a long time ago, found that this could be obtained by introducing a duality $ds \to l_0^2/ds$ in the particle path integral \cite{Padmanabhan:1996ap,Padmanabhan:1998yya}. With this, ultraviolet contributions are suppressed and,  interestingly enough, it is equivalent to a change in the spacetime interval given by $ds^2 \to ds^2 + l_0^2$.

This determines the introduction of a ``zero-point length'', $l_{0}$, in the field propagator, whose momentum space representation reads
\begin{equation}
G(\vec{k}) = - \frac{l_{0}}{\sqrt{k^{2}+m^{2}}}
K_{1}\left(l_{0}\sqrt{k^{2}+m^{2}}\right),
\label{propagator}
\end{equation}
where $K_{1}(z)$ is a modified Bessel function of the second kind. The idea of introducing a duality  such as $ds\to L^2/ds$ has some similarities with string theory, where it is shown that small-radius torus compactifications are dual to large-radius compactifications, and this is called $T-$duality. In fact, some time latter, it was shown that the above particle duality can emerge form string theory \cite{Smailagic:2003hm,Spallucci:2005bm,Fontanini:2005ik}.

However, just very recently, the new propagator (\ref{propagator}) has been used to compute the effective potential. Furthermore, one can show that the potential follows to be \cite{Nicolini:2019irw}
\begin{align}
V(x) =&-M \int \frac{d^{3}k}{(2\pi)^{3}} G_{F}(k)\big|_{k^{0}=0} \exp(i\vec{k}\cdot\vec{x})
\\
 =& -\frac{M}{\sqrt{x^{2}+l_{0}^{2}}}.
\end{align}
induced by the path integral duality or, equivalently, by T-duality. The expression for the potential can be naturally understood as describing a smeared matter distribution, rather than a conventional pointlike source. In this picture, the source is effectively regularized at short distances. For a review, see reference \cite{Nicolini:2022rlz}.

The energy density associated with the modified matter distribution follows directly from the Poisson equation, giving  
\begin{equation}  
\rho(x) = \frac{1}{4\pi}\,\Delta V(x) = \frac{3 l_{0}^{2} M}{4\pi (x^{2}+l_{0}^{2})^{5/2}}.  
\label{density}  
\end{equation}  
This expression provides a natural framework to investigate quantum-induced modifications of spacetime geometry, illustrating how nonlocal effects arise as a direct consequence of the presence of a minimal length. Using Eq.~\eqref{density}, and considering a spherically symmetric black hole metric written as
\begin{equation}
ds^{2} = -g_{00}(x) dt^{2} + g_{11}(x)dx^{2} + x^{2}\, d\Omega^{2},
\end{equation}
one obtains the following $T$-duality inspired black hole solution:  
\begin{equation}  
-g_{00}(x) = g_{11}^{-1}(x) = 1 - \frac{2 M x^{2}}{(x^{2}+l_{0}^{2})^{3/2}}.  
\label{metric}  
\end{equation}  
Interestingly, this metric coincides in form with the well-known Bardeen black hole \cite{2767662}. Shortly thereafter, the ideas developed in Refs.~\cite{Padmanabhan:1996ap,Nicolini:2019irw} were applied to construct finite electrodynamics models and regular charged black holes \cite{Gaete:2022une,Gaete:2022ukm}. More recently, several aspects of these geometries have been analyzed, including quasi-normal modes, thin-shell dynamics, geodesic completeness, and wormholes \cite{Konoplya:2023ahd,Javed:2024wbc,Jusufi:2024dtr,Lobo:2025nng}.  

However, there is a class of spacetimes that has not been considered in the literature. After its proposal, black bounces (BBs) have been widely studied and have attracted considerable interest among researchers in gravitation due to their intriguing geometric structure, in which the black hole singularity is removed and replaced by a wormhole throat \cite{Bolokhov:2024sdy}.

The first known BB metric appeared in the literature in 1997 in a work by Hochberg and Visser \cite{Visser:1997yn}, and was later studied in detail within a modern framework by Simpson and Visser in 2018 \cite{Simpson:2018tsi}. Subsequently, several BB models have been proposed \cite{Lobo:2020ffi,Simpson:2019cer,Franzin:2021vnj,Furtado:2022tnb,Lima:2022pvc,Lima:2023jtl,Crispim:2024nou,Crispim:2024yjz,Alencar:2025jvl,Rodrigues:2022mdm,Rodrigues:2022rfj,Pereira:2023lck,Rodrigues:2025plw,Bronnikov:2023aya,Lima:2023arg,Alencar:2024yvh,Alencar:2024nxi}, considering different symmetries, contexts, and construction methods. In addition to being free of singularities, many BBs can even exhibit observational features similar to their singular counterparts \cite{Guerrero:2021ues, Lima:2021las,Vagnozzi:2022moj, Tsukamoto:2020bjm}, while other models display observational characteristics that differ from those of classical black holes \cite{Crispim:2025cql,Arora:2023ltv,Duran-Cabaces:2025sly}. This makes them particularly relevant in the context of testing gravitational theories and confronting astrophysical observations.

One of the disadvantages of BBs is the fact that, in general, the source is given by a non-linear electrodynamics (NED) \cite{Alencar:2024yvh,Alencar:2025jvl,Bronnikov:2021uta,Canate:2022gpy,Rodrigues:2023vtm}. Using NED as the matter source for BB spacetimes is attractive because it yields regular geometries, but it brings concerns: Photon propagation is altered in such a way that they do not necessarily follow the light cones of the background metric $g_{\mu\nu}$. In nonlinear regimes, photons instead follow the null geodesics of an effective metric determined by the background field, leading to the phenomenon of birefringence, where different polarizations experience distinct light cones \cite{Novello:1999pg,Russo:2022qvz,deMelo:2014isa}. Some NED models also can exhibit acausal or ill-posed behavior in the presence of strong fields. Specifically, in regions where the field strengths approach or exceed the characteristic scales of the theory, the effective metric governing photon propagation can develop pathological features, such as superluminal modes or a breakdown of hyperbolicity. These issues imply that the evolution of small perturbations may not be uniquely determined by the initial data, signaling a loss of predictability and potential violations of causality \cite{Russo:2024llm,deMelo:2014isa,Abalos:2015gha,dePaula:2024yzy}. Consequently, care must be taken when interpreting solutions in strong-field regimes, as certain NED models may not provide a physically consistent description under these conditions.

In this manuscript, we study the consequences of replacing the NED by the quantum corrected string source \eqref{density}. In this case, we adopt general relativity as our gravitational theory and, starting from a quantum source motivated by T-duality, we investigate whether it can give rise to a BB geometry. This approach provides, among other aspects, a natural physical interpretation for the bounce regularization parameter, which is here understood as a quantum correction induced by T-duality. This work is therefore organized as follows: In section \ref{S:Fieldeq}, we will present and solve the field equations considering this regularized source, as well as analyze the main physical properties of our solution. In the section \ref{S:SV}, we perform a detailed analysis of the geodesic motion for both massive and massless particles, constraining the minimal length parameter through the black hole shadow radius obtained from EHT data and examining the optical appearance of the spacetime with a thin accretion disk. We then investigate in the section \ref{S:HCS} the thermodynamic stability of the solution via the heat capacity, identifying possible phase transitions, and analyze the energy conditions in section \ref{S:conclusion}, highlighting those violated by the effective fluid supporting the geometry. The section summarizes our main results and outlines prospects for future research.


 We adopt the metric signature $(-,+,+,+)$.
 We shall work in geometrodynamics units where $ G=\hbar=c=1$. 

\section{Black bounce solution}
\label{S:Fieldeq}

\subsection{Field equations}
To find BB solutions, we consider the line element written as
\begin{equation}\label{line_element}
ds^2= -A\left(x\right)dt^2+ \frac{dx^2}{A\left(x\right)} + \Sigma\left(x\right)^2d\Omega_2^2,
\end{equation}
where $A(x)$ and $\Sigma(x)$ are general functions of the radial coordinate and $d\Omega_2^2 = d\theta^2 + \sin^2\theta  d\varphi^2$ is the 2-sphere line element.

To find the metric coefficients, we consider general relativity as our gravitational theory, so that the functions can be obtained from the Einstein equations given by
\begin{eqnarray}
    G_{\mu\nu}=R_{\mu\nu}-\frac{1}{2}Rg_{\mu\nu}=\kappa^2 T_{\mu\nu},\label{einstein}
\end{eqnarray}
where $G_{\mu\nu}$ is the Einstein tensor, $T_{\mu\nu}$ is the stress-energy tensor, $R_{\mu\nu}$ is the Ricci tensor, $R$ is the curvature scalar, $g_{\mu\nu}$ is the metric tensor, and $\kappa^2=8\pi$.

We can consider the stress-energy tensor for an anisotropic fluid, which is written as
\begin{equation}
    T_{\mu\nu} = (\rho + p_t)\, u_\mu u_\nu + p_t\, g_{\mu\nu} + (p_r - p_t)\, \chi_\mu \chi_\nu,
\end{equation}
with $u^\mu$ and $\chi^\mu$ being the normalized timelike and spacelike vectors, respectively, and we have $u^\mu u_\mu = -1$, $\chi^\mu \chi_\mu = 1$, and $u^\mu \chi_\mu = 0$. Furthermore, we can identify $\rho$ as the energy density, $p_r$ as the radial pressure, and $p_t$ as the tangential pressure.

In comoving coordinates, we can simplify the form of the stress-energy tensor by writing it as
\begin{equation}
    T^{\mu}{}_{\nu}={\rm diag}\left[-\rho,\,p_r,\,p_t,\,p_t\right].\label{SE_tensor}
\end{equation}

Considering the line element \eqref{line_element} together with the components of the stress-energy tensor \eqref{SE_tensor}, the components of the Einstein equations are given by
\begin{eqnarray}
    \frac{A' \Sigma '}{\Sigma }+\frac{2 A \Sigma ''}{\Sigma }+\frac{A \Sigma '^2}{\Sigma
   ^2}-\frac{1}{\Sigma ^2}=-\kappa^2 \rho,\label{Einstein1}\\
   \frac{A' \Sigma '}{\Sigma}+\frac{A \Sigma '^2}{\Sigma ^2}-\frac{1}{\Sigma ^2}=\kappa^2 p_r,\label{Einstein2}\\
   \frac{A''}{2}+\frac{A' \Sigma '}{\Sigma }+\frac{A \Sigma ''}{\Sigma }=\kappa^2p_t,\label{Einstein3}
\end{eqnarray}
where {${}'$} denotes the derivative with respect to $x$. In addition to the components of Einstein's equations, we also have the conservation of the stress-energy tensor, which is given by $\nabla_\mu T^{\mu\nu} = 0$. This condition leads to
\begin{equation}
    \frac{1}{2} A' (p_r+\rho) +A p_r'+\frac{2 A (p_r-p_t) \Sigma
   '}{\Sigma }=0.\label{conservation}
\end{equation}
Equations \eqref{einstein}–\eqref{conservation} will be used to determine the BB solution.

To verify the regularity of this solution, we must analyze the Kretschmann scalar, which is given by
\begin{equation}
\begin{split}
K = \frac{1}{\Sigma^4} \Big[ (\Sigma^2 A'')^2 + 2(\Sigma A' \Sigma')^2 + 2\Sigma^2 (A' \Sigma' + 2A \Sigma'')^2 \\
+ 4(1 - A \Sigma^2)^2 \Big].
\end{split}
\end{equation}
To ensure the regularity of the spacetime, we must have \cite{Lobo:2020ffi}
\begin{itemize}
    \item $\Sigma \neq 0$ everywhere;
    \item  $\Sigma'$ and $\Sigma''$ finite everywhere;
    \item $A$, $A'$, and $A''$ finite everywhere.
\end{itemize}

We can also analyze the behavior of the Hernandez-Misner-Sharp quasilocal mass, given by
\begin{equation}
    M_{\text{HMS}}(x) = \frac{1}{2} \Sigma \left( 1 - A\Sigma'^2 \right).\label{MHMS}
\end{equation}
\subsection{Black bounce from string T-duality}

The energy density obtained from the obtained from the string T-duality is given by
\begin{equation}
    \rho = \frac{3M l_0^2}{4\pi(x^2 + l_0^2)^{5/2}}.\label{dens}
\end{equation}
Consequently, we need to determine the functions $p_r$, $p_t $, $ A $, and $ \Sigma $. As is common in BB solutions with standard spherical symmetry, we assume 
\begin{equation}\label{sigma}
    \Sigma(x)^2 = x^2 + l_0^2,
\end{equation}
so that the area of the symmetry 2-spheres is given by $ A_{S^2} = 4\pi \Sigma^2 = 4\pi(x^2 + l_0^2) $, and satisfies all the conditions that $\Sigma$ must fulfill in order for the Kretschmann scalar to remain regular throughout the entire manifold. This area attains a minimum at $ x = 0 $, where $ A_{S^2} = 4\pi l_0^2 $. Thus, we are imposing that the size of the bounce throat is determined by the constant $ l_0 $. From a mathematical standpoint, this choice is reasonable since the function $ \rho(x) $, which governs the structure of the metric component $ A(x) $, remains regular at the origin due to the presence of $ l_0 $. Physically, such fact might be associated with the singularity removal via quantum corrections induced by the energy density arising from string T-duality. 

Once the function $ \Sigma(x) $ is fixed, the metric coefficient $ A(x) $ can be obtained from \eqref{Einstein1}, which is given by
\begin{equation}
    A(x)=1+\frac{C_1 \sqrt{l_0^2+x^2}}{x^2}+\frac{3 l_0^2 M}{x^2 \sqrt{l_0^2+x^2}}+\frac{l_0^2}{x^2},\label{Ax1}
\end{equation}
where $C_1$ is an integration constant. 

As mentioned earlier, in order to guarantee the regularity of the solution, we require that $A(x)$ remains finite. However, we can see that equation \eqref{Ax1} is not finite as $x \to 0$. Therefore, we can fix the integration constant by imposing that $A(x)$ remains finite in the limit $x \to 0$, which leads to the choice $ C_1 = -3M - l_0 $. With this, we have
\begin{equation}\label{Ax}
    A(x) =1+ \frac{3 l_0^2 M}{x^2 \sqrt{l_0^2+x^2}}-\frac{(l_0+3 M) \sqrt{l_0^2+x^2}}{x^2}+\frac{l_0^2}{x^2}.
\end{equation}

In addition to obtaining the metric functions, we also need to determine the form of the remaining components of the stress-energy tensor. For this purpose, we can use the equations \eqref{Einstein2} and \eqref{conservation}. From these we obtain the radial and tangential pressures, given by
\begin{eqnarray}
    p_r&=&\frac{l_0^2 \left(l_0-\sqrt{l_0^2+x^2}\right)}{4 \pi  x^2 \left(l_0^2+x^2\right)^{3/2}},\\
    p_t&=&\frac{l_0^2\left(3 l_0^2+2 x^2\right)}{8 \pi x^4\left(  l_0^2 +  x^2\right)}\nonumber\\
    &&-\frac{l_0^2 \left(6 l_0^5+13 l_0^3 x^2+x^4 (7 l_0+3
   M)\right)}{16 \pi  x^4 \left(l_0^2+x^2\right)^{5/2}}.
\end{eqnarray}
Thus, we see that the source of this solution is an anisotropic fluid.

The Hernandez–Misner–Sharp mass, \eqref{MHMS}, for our solution is given by
\begin{equation}
    M_{HMS}=\frac{l_0^3+l_0 x^2+3 M x^2}{2 \left(l_0^2+x^2\right)}.\label{MHMS1}
\end{equation}
We see that the expression is always positive and has the following limits
\begin{equation}
    \lim_{x\to 0} M_{HMS}=\frac{l_0}{2}, \quad \mbox{and} \quad \lim_{x\to \infty} M_{HMS}=\frac{1}{2} (l_0+3 M).
\end{equation}
If we define the ADM mass as $M_{ADM}=\lim_{x \to \infty} M_{\text{HMS}}$, then the asymptotic behavior of our solution can be rewritten as
\begin{equation}\label{Alimite}
    A(x \to \infty) \approx 1 - \frac{2 M_{ADM}}{x} + O(1/x^2),
\end{equation}
where the ADM mass receives contributions from both $M$ and $l_0$. 

\subsection{Causal structure and Carter-Penrose diagrams}
Our resulting metric is then defined as \eqref{line_element}, where the metric functions $\Sigma(x)$ and $A(x)$ are given respectively by \eqref{sigma} and \eqref{Ax} and the coordinates take values within the intervals
\begin{equation}
    t \in (-\infty, + \infty), \; x \in (-\infty, + \infty), \; \theta \in [0, \pi], \; \phi \in [0, 2\pi]. 
\end{equation}

The asymptotic forms of $A(x)$ are given by
\begin{eqnarray}
    A(x \to 0) &\approx& \frac{1}{2} - \frac{3M}{l_0} + O(x^2),\\
    A(x \to \infty) &\approx& 1 - \frac{(3M +l_0)}{x} + \frac{l_0^2}{x^2} + O(1/x^3).  
\end{eqnarray}
With this, we find that the solution is asymptotically flat and everywhere finite.

The analytical expression for the Kretschmann scalar $K = R_{\alpha\beta\mu\nu}R^{\alpha\beta\mu\nu}$ is quite lengthy, but we can analyze its asymptotic behavior, which is given by
\begin{eqnarray}
    K(x \to 0) &\approx& \frac{97 l_0^2-360 l_0 M+1296 M^2}{16 l_0^6}+ O(x^2),\nonumber\\ \\
    K(x\to \infty) &\approx& \frac{12 (l_0+3 M)^2}{x^6} +O(1/x^7).
\end{eqnarray}

Solving $A(x)=0$, we find the radius of the event horizon, which is given by
\begin{equation}\label{xh}
    x_{H\pm}=\pm\frac{\sqrt{-2 l_0^2+3 M^{3/2} \sqrt{12 l_0+9 M}+6 l_0 M+9 M^2}}{\sqrt{2}}.
\end{equation}
For $l_0 > 6M$, there are no event horizons. In terms of the ADM mass, we have no horizons to $l_0>4M_{ADM}/3$. Thus, we obtain a solution  regular and asymptotically flat that resembles the Simpson-Visser model: 
\begin{itemize}
\item For $l_0> 6M$, there are no event horizons, and we thus have a traversable wormhole;
    \item For $l_0 = 6M$, we have a horizon located at $x=0$, which shows that this solution corresponds to a wormhole with a null extremal throat located at $x=0$;
    \item For $l_0 < 6M$, we find that the throat lies inside the horizon, and therefore the solution represents a regular black hole with the pair of horizons given by \eqref{xh}.
\end{itemize}

\begin{figure}[htb]
    \centering
    \includegraphics[width=1.\linewidth]{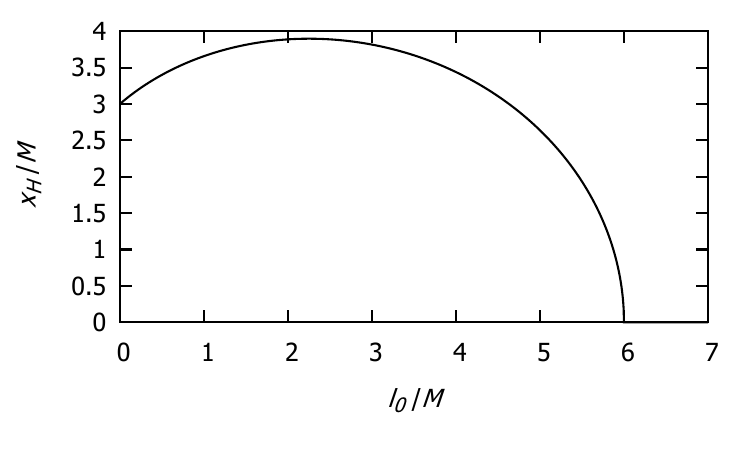}
    \caption{Event horizon radius as a function of $l_0$.}
    \label{fig:xh}
\end{figure}

In Fig. \ref{fig:xh}, we plot the radius of the event horizon as a function of the parameter $l_0$, where one can visually confirm the behavior previously discussed: An event horizon exists only for values of $l_0/M <6$. Furthermore, it can be seen that there is a specific value of $l_0$ for which the event horizon reaches its maximum size. This particular value is given by $l_0/M = 9/4$, which corresponds to a maximum horizon radius equal to ${x_H}^{max}/M = 9\sqrt{3}/4$.

\begin{figure}[htb]
    \centering
    \includegraphics[width=.85\linewidth]{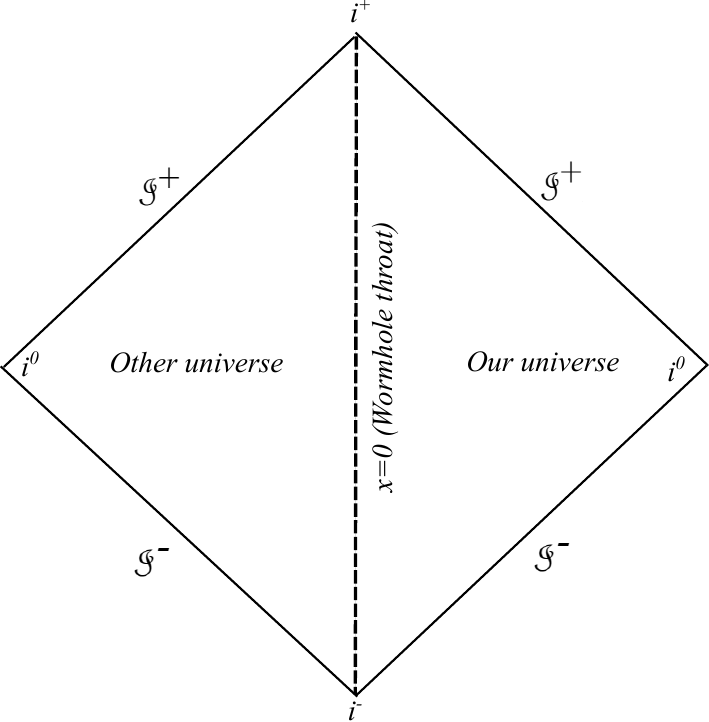}
    \caption{Carter-Penrose diagram for the case $l_0 >6M$, where we have a traversable wormhole.}
    \label{fig:diagramTW}
\end{figure}

To better understand the causal structure of this solution and how it changes with variation of the parameter $l_0$, we can use the Carter-Penrose diagrams. In the case of $l_0 > 6M$, when we have the transversable wormhole, we must note that the throat is a timelike hypersurface located at $x=0$, where the negative values of the radial coordinate correspond to the universe on the other side of the geometry, from the point of view of an observer located at positive values of the radial coordinate. Both universes are separated by the throat at $x=0$. Therefore, the Carter-Penrose diagram for this case is simply the usual diagram for a traversable wormhole geometry, as represented in Fig.~\ref{fig:diagramTW}.

\begin{figure}[htb]
    \centering
    \includegraphics[width=.7\linewidth]{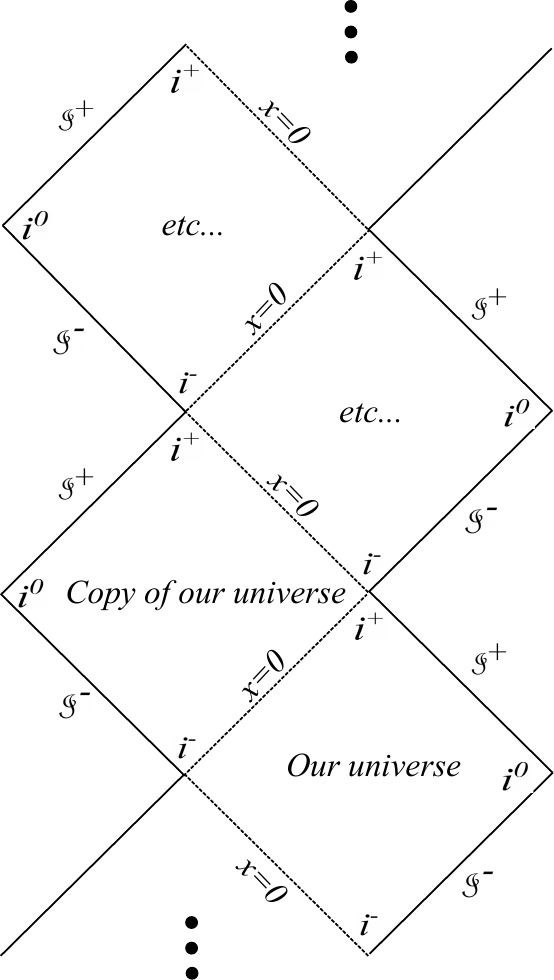}
    \caption{Carter-Penrose diagram for the case $l_{0}=6M$, where the null throat coincides with the extremal horizon. The geometry is one way traversable, and the maximal analytic extension produces an infinite chain of asymptotic regions.}
    \label{fig:oneway}
\end{figure}

Similarly to the null case of the Simpson-Visser model, when the parameter reaches the critical value $l_{0} = 6M$, the throat becomes null and coincides with the horizon located at $x = 0$. In this situation, the root of $A(x)$ at $x=0$ is degenerate (extremal horizon), which implies vanishing surface gravity and a peculiar causal behavior: the geometry admits passage through the throat only in one direction (one way traversable), so that geodesics crossing the horizon or throat do not return to the same asymptotic region. The Carter-Penrose diagram of the maximally extended spacetime reflects this structure, where the null throat appears superimposed with the extremal horizon, and the analytic extension produces a single type of connection between the asymptotic sectors, as illustrated in Fig.~\ref{fig:oneway}.

\begin{figure}[htb]
    \centering
    \includegraphics[width=.83\linewidth]{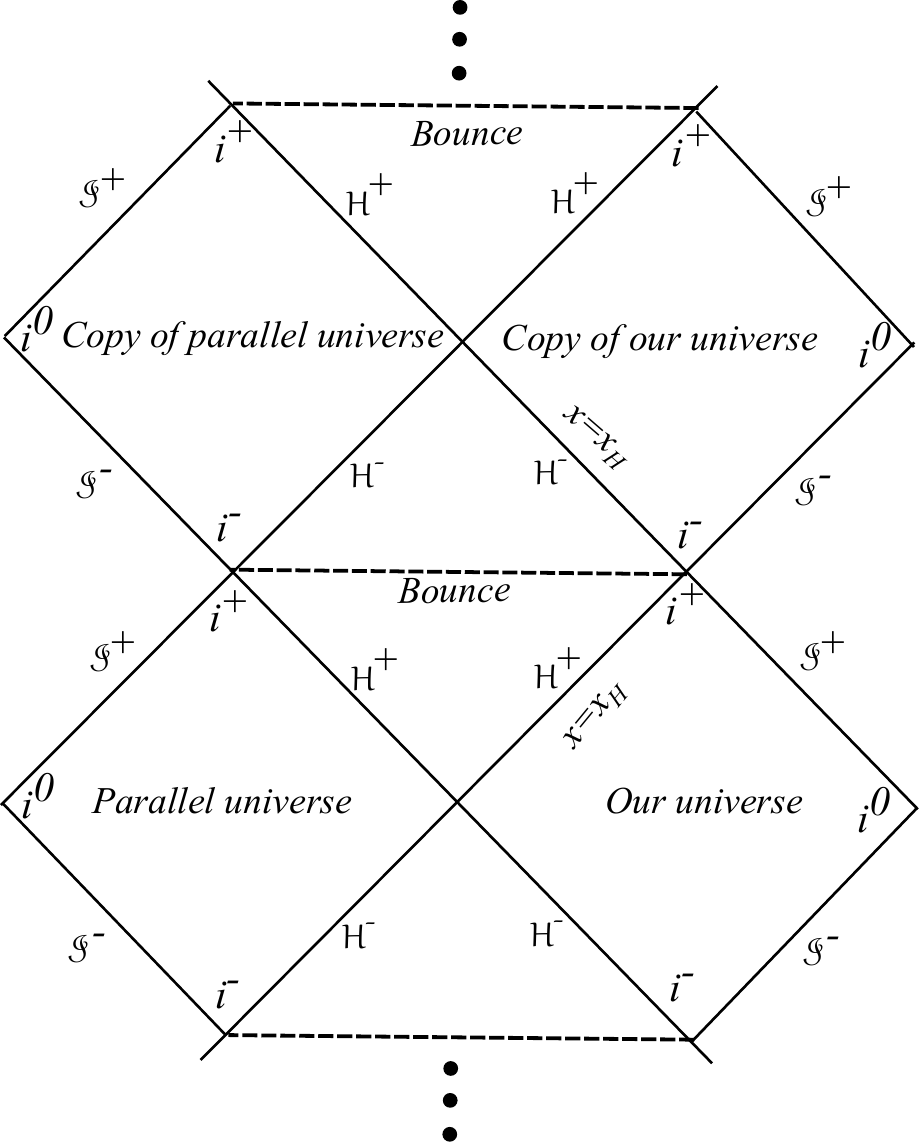}
    \caption{Carter-Penrose diagram for the case $l_{0}<6M$, where the throat is hidden behind the horizon and the singularity is replaced by a spacelike bounce.}
    \label{fig:regular}
\end{figure}

For the case of regular black holes, obtained when $l_{0}<6M$, the throat is hidden behind the event horizon. The surface at $x=0$ is now spacelike and acts as a bounce that links our universe to another copy of itself. In this scenario, the Schwarzschild singularity is avoided and replaced by a spacelike boundary. 
Crossing to negative values of $x$ corresponds to entering a new asymptotic region, as shown in Fig.~\ref{fig:regular}. 
This causal structure differs from the maximally extended Kruskal-Szekeres diagram of Schwarzschild, 
where a timelike singularity is present. 
Here, instead, the interior geometry undergoes a bounce that provides a smooth continuation into another universe.
(See, for example, Refs.~\cite{Penrose1965,HawkingEllis1973,Carter1968} for classical analyses of Carter-Penrose diagrams.)

Our causal structure closely resembles that of the Simpson–Visser model. Next, we will study the trajectories of massive and massless particles in our geometry.


\section{Geodesics and shadow radius}\label{S:SV}

A standard approach to probing the structure of a given spacetime is through the analysis of particle motion within that geometry, namely, by studying its geodesics. This framework allows for the extraction of several physical observables, such as the optical appearance and the shadow cast by the object. The geodesic motion is defined as $ds^2/d\lambda^2 = \epsilon$, where $\epsilon =-1$ for massive particles and $\epsilon=0$ for massless particles. Given our metric, this leads to
\begin{equation}\label{geodesic1}
    -A\dot{t}^2 + \frac{\dot{x}^2}{A} + \Sigma^2\dot{\theta}^2 + \Sigma^2\sin^2\theta\dot{\phi}^2 = \epsilon,
\end{equation}
where the dot is the derivative with respect to the affine parameter $\lambda$. Taking into account that this metric possesses a timelike Killing vector associated with energy conservation and a rotational Killing vector associated with the conservation of angular momentum in the $z$-direction, we also have the following equations
\begin{equation}\label{killing}
    A\dot{t} = E, \;\;\;\; \Sigma^2\dot{\phi} = \ell,
\end{equation}
which, after substituting into Eq. \eqref{geodesic1} and taking $\theta=\pi/2$ (equatorial plane), yields the equation for the coordinate $x$
\begin{equation}\label{xeq}
    \dot{x}^2 = E^2 + A\left(\epsilon - \frac{\ell^2}{\Sigma^2}\right).
\end{equation}

We can now separate our study into the cases of null and timelike geodesics.

\subsection{Time-like geodesics}
Timelike geodesics are defined for $\epsilon=-1$. Thus, we can rewrite equation \eqref{xeq} as
\begin{equation}
    \dot{x}^2 = E^2 - V_m(x),
\end{equation}
where
\begin{equation}
    V_m(x) = A\left(1 + \frac{\ell^2}{\Sigma^2}\right)
\end{equation}
is the effective potential for massive particles. Explicitly, this is given by
\begin{align}
V_m(x) &=   \left(\frac{\ell^2}{l_0^2+x^2}+1\right) \times \notag \\
& \left(\frac{3 l_0^2 M}{x^2 \sqrt{l_0^2+x^2}}-\frac{(l_0+3 M) \sqrt{l_0^2+x^2}}{x^2}+\frac{l_0^2}{x^2}+1\right).
\end{align}\label{Vmass}

The so-called innermost stable circular orbits (ISCO) are defined in terms of the effective potential for massive particles. In general, circular orbits at $x=x_c$ satisfy $V_m'(x_c)=0$, while stability requires $V_m''(x_c)>0$ and instability $V_m''(x_c) <0$. The ISCO corresponds to the limiting case where stability is lost, namely $V_m'(x_{\rm ISCO})=0$ and $V_m''(x_{\rm ISCO})=0$. Analytical expressions for $x_{\rm ISCO}$ cannot be obtained for the potential given by Eq.\eqref{Vmass}; however, they can be computed numerically.

\begin{figure*}[htb]
    \centering
        \includegraphics[width=.5\linewidth]{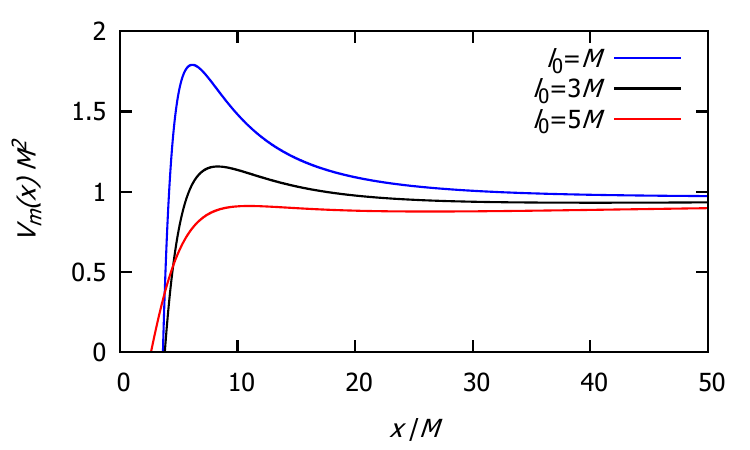}\hspace{-0.1cm}
    \includegraphics[width=.5\linewidth]{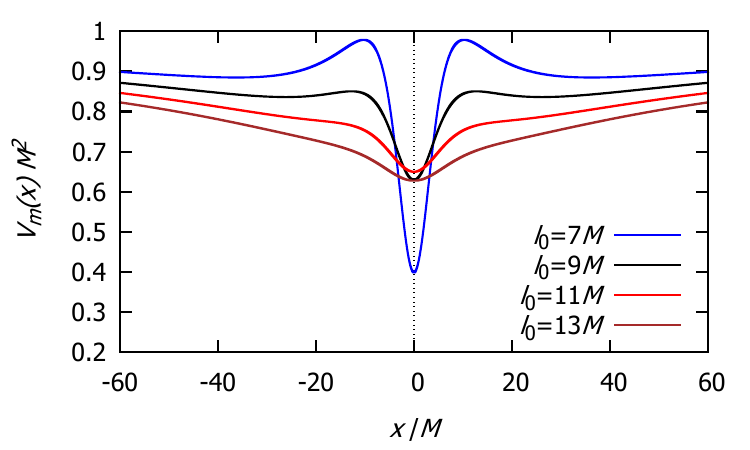}
    \caption{Effective potential for massive particles as a function of the radial coordinate for different values of $l_0$, for the case with  horizon fixing $\ell = 12M$ (top panel) and the case without a horizon fixing $\ell = 15M$ (bottom panel).}
    \label{fig:Vmass}
\end{figure*}

In Fig.~\ref{fig:Vmass} we show the plot of the effective potential for massive particles as a function of $x$ for different values of the parameters. Although the existence of stable orbits is not clearly visible in the plot, it can be obtained numerically for the values considered. In Table~\ref{tab} we present the numerical values of the position of the stable ($x_{sc}$) and unstable ($x_{uc}$) orbits considering the parameter values used in the plot of Fig.~\ref{fig:Vmass}. We can observe that, for the values considered, all values of $l_0$ admit an orbit in the case of a regular black hole, which is not the case for the wormhole, where for the values of $l_0$ given by $l_0 =11M$ and $l_0 =13M$ the unstable orbit does not exist.

\begin{widetext}
\centering
\begin{table*}[!htb]
\begin{tabular}{lllllllll}
\cline{1-8}
\multicolumn{1}{c|}{}             & \multicolumn{3}{c|}{Regular black hole (fixing $\ell =12M$)}                                                                  & \multicolumn{4}{c}{Wormhole (fixing $\ell=15M$)}                                                                                               &                      \\ \cline{1-8}
\multicolumn{1}{c|}{$l_0$}             & \multicolumn{1}{c|}{$M$}          & \multicolumn{1}{c|}{$3M$}        & \multicolumn{1}{c|}{$5M$}         & \multicolumn{1}{c|}{$7M$}         & \multicolumn{1}{c|}{$9M$}         & \multicolumn{1}{c|}{$11M$} & \multicolumn{1}{c}{$13M$} &                      \\ \cline{1-8}
\multicolumn{1}{c|}{$x_{\rm sc} \approx$} & \multicolumn{1}{c|}{$ 65.96M$} & \multicolumn{1}{c|}{$ 40.40M$} & \multicolumn{1}{c|}{$ 26.25M$} & \multicolumn{1}{c|}{$ 36.23M$ and $0$} & \multicolumn{1}{c|}{$26.14M$ and $0$} & \multicolumn{1}{c|}{$0$}       & \multicolumn{1}{c}{$0$}       &                      \\ \cline{1-8}
\multicolumn{1}{c|}{$x_{\rm uc} \approx$} & \multicolumn{1}{c|}{$ 6.15M$}  & \multicolumn{1}{c|}{$ 8.23M$}  & \multicolumn{1}{c|}{$ 10.97M$} & \multicolumn{1}{c|}{$ 10.25M$} & \multicolumn{1}{c|}{$ 12.72M$} & \multicolumn{1}{c|}{DNE}       & \multicolumn{1}{c}{DNE}       & \multicolumn{1}{c}{} \\ \cline{1-8}
\end{tabular}
\caption{Stable and unstable orbits positions for the black hole ($l_0 <6M$) and wormhole ($l_0 >6M$) cases, for different values of $l_0$. The abbreviation DNE stands for "Does Not Exist," which means that for these values no orbit exists.}
\end{table*}\label{tab}
\end{widetext}

In Table \ref{tabisco} we have the values of the ISCO positions for different values of the parameter $l_0$, as well as the values of angular momentum associated with each ISCO.

\begin{widetext}
\centering
\begin{table*}[htb]
\begin{tabular}{lllllllll}
\cline{1-8}
\multicolumn{1}{c|}{}                  & \multicolumn{3}{c|}{Regular black hole}                                                                                          & \multicolumn{4}{c}{Wormhole}                                                                                                                                                    &                      \\ \cline{1-8}
\multicolumn{1}{c|}{}                  & \multicolumn{1}{c|}{$l_0=M$}           & \multicolumn{1}{c|}{$l_0 =3M$}         & \multicolumn{1}{c|}{$l_0=5M$}          & \multicolumn{1}{c|}{$l_0=7M$}          & \multicolumn{1}{c|}{$l_0=9M$}          & \multicolumn{1}{c|}{$l_0=11M$}         & \multicolumn{1}{c}{$l_0=13M$}                        &                      \\ \cline{1-8}
\multicolumn{1}{c|}{$x_{\rm ISCO}$}    & \multicolumn{1}{c|}{$\approx 11.299M$} & \multicolumn{1}{c|}{$\approx 14.022M$} & \multicolumn{1}{c|}{$\approx 15.724M$} & \multicolumn{1}{c|}{$\approx 16.877M$} & \multicolumn{1}{c|}{$\approx 17.643M$} & \multicolumn{1}{c|}{$\approx 18.100M$} & \multicolumn{1}{c}{$\approx 18.285M$}                &                      \\ \cline{1-8}
\multicolumn{1}{c|}{$\ell_{\rm ISCO}$} & \multicolumn{1}{c|}{$\approx 6.704M$}  & \multicolumn{1}{c|}{$\approx 9.056M$}  & \multicolumn{1}{c|}{$\approx 10.998M$} & \multicolumn{1}{c|}{$\approx 12.710M$} & \multicolumn{1}{c|}{$\approx 14.267M$} & \multicolumn{1}{c|}{$\approx 15.709M$} & \multicolumn{1}{c}{$\approx 17.061M$} & \multicolumn{1}{c}{} \\ \cline{1-8}          
\end{tabular}
\caption{ISCO positions for the black hole ($l_0 <6M$) and wormhole ($l_0 >6M$) cases, for different values of $l_0$. The abbreviation DNE stands for "Does Not Exist," meaning that for these values no orbit exists.}
\end{table*}\label{tabisco}
\end{widetext}

\begin{figure*}[htb]
    \centering
    \includegraphics[width=.5\linewidth]{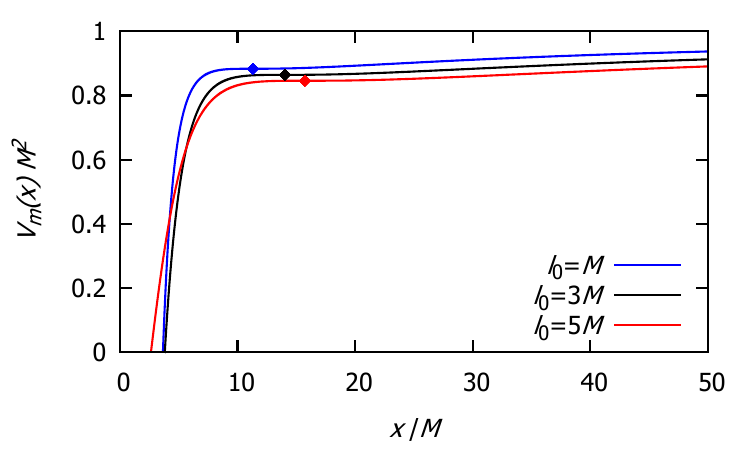}\hspace{-.1cm}
    \includegraphics[width=.5\linewidth]{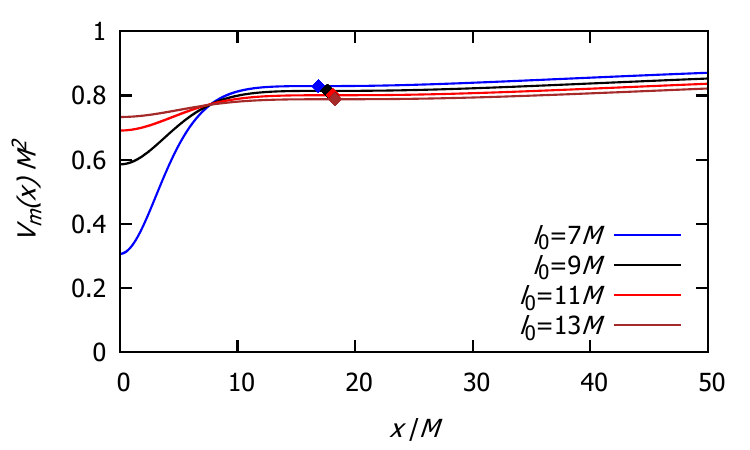}
    \caption{The plots show the effective potential and the positions of the ISCO orbits for the case of a regular black hole (top panel) for the values of $l_0$ given by $l_0 = \{M, 3M, 5M\}$, whose corresponding angular momentum values associated with the ISCO orbits are $\ell_{\rm ISCO} \approx \{6.704M, 9.056M, 10.998M\}$, and for the case of a wormhole (bottom panel) for the values of $l_0$ given by $l_0 = \{7M, 9M, 11M, 13M\}$, whose corresponding angular momentum values associated with the ISCO orbits are $\ell_{\rm ISCO} = \{12.710M, 14.267M, 15.709M, 17.061M\}$.
}
    \label{fig:VISCO}
\end{figure*}

In Fig. \ref{fig:VISCO} we have the plot of the effective potential for massive particles considering the $l_0$ values from the previous table and the angular momentum $\ell=\ell_{ISCO}$. Each curve shows the marked position of the associated ISCO position.

\begin{figure}[htb]
    \centering
    \includegraphics[width=1\linewidth]{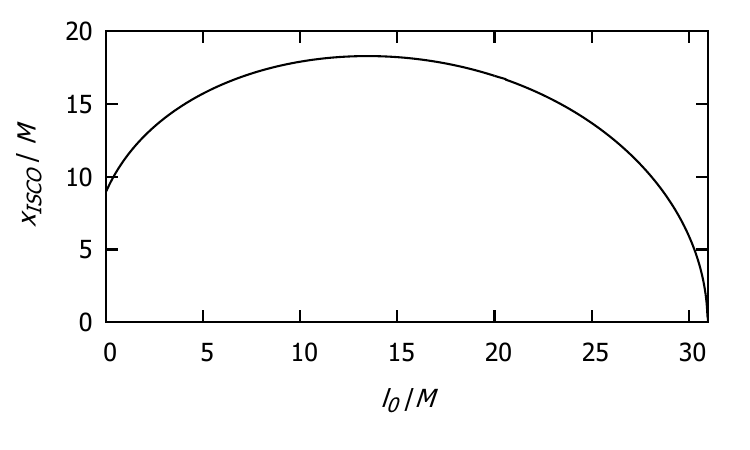}
    \caption{ISCO orbit position as a function of the parameter $l_0$.}
    \label{fig:xisco}
\end{figure}

\begin{figure}[htb]
    \centering
    \includegraphics[width=1\linewidth]{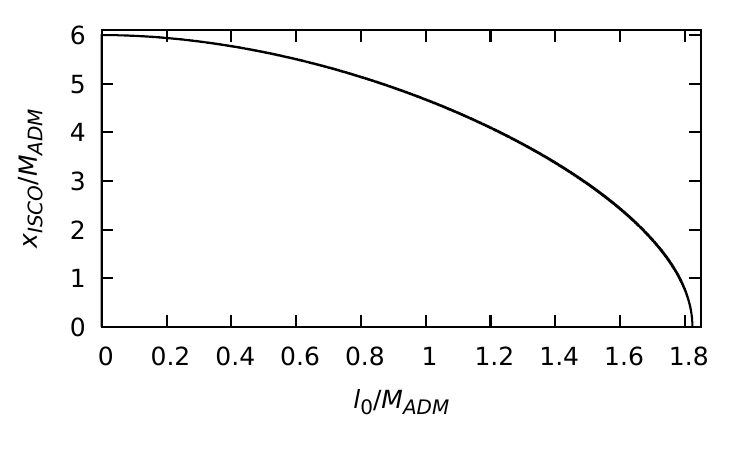}
    \caption{ISCO position as a function of the parameter $l_0$ when we normalize in terms of the AMD mass.}
    \label{fig:xisco_adm}
\end{figure}

An interesting point to analyze is how the position of the ISCO orbit varies as a function of the parameter $l_0$. In Fig.~\ref{fig:xisco} we show $x_{\text{ISCO}}$ as a function of $l_0$. One can see that $x_{\text{ISCO}}$ initially increases with $l_0$, reaching a maximum at approximately $l_0/M \approx 13.75$, where the corresponding orbit is located at $x_{\text{ISCO}}^{\text{max}}/M \approx 18.29$. After this point, as $l_0$ increases further, $x_{\text{ISCO}}$ decreases until it vanishes at $l_0/M \approx 31.05$. We can also analyze the behavior of $x_{\text{ISCO}}$ in terms of $l_0$, taking the ADM mass into account. Since our solution behaves asymptotically like Schwarzschild when the ADM mass is considered, it is natural to expect that, in some limit, $x_{\text{ISCO}}$ will approach the Schwarzschild value. Figure \ref{fig:xisco_adm} confirms this behavior: as $l_0 \to 0$, the value of $x_{\text{ISCO}}$ approaches $6M_{\text{ADM}}$. As we shall see later, these results are of fundamental importance when modeling accretion disks and studying the optical appearance of the object.

\subsection{Null geodesics}

 In this case, we take $\epsilon=0$ in equation \eqref{xeq}. By absorbing $\ell$ into the affine parameter, the equation can further be rewritten as
\begin{equation}
    \dot{x}^2 = \frac{1}{b^2} - V_\gamma(x),\label{radial_velocity}
\end{equation}
where $b \equiv \ell/E$ is the impact parameter and $V_\gamma(x)$ is the effective potential for photons given by
\begin{equation}
    V_\gamma(x) = \frac{A}{\Sigma^2},\label{V_gen}
\end{equation}
or, explicitly
\begin{equation}
    V_\gamma(x) = \frac{1}{x^2}\left(1 - \frac{l_0}{\sqrt{x^2 + l_0^2}}\right) - \frac{3M}{ (x^2 + l_0^2)^{3/2}}.
\end{equation}

\begin{figure*}[htb]
    \centering
    \includegraphics[width=.5\linewidth]{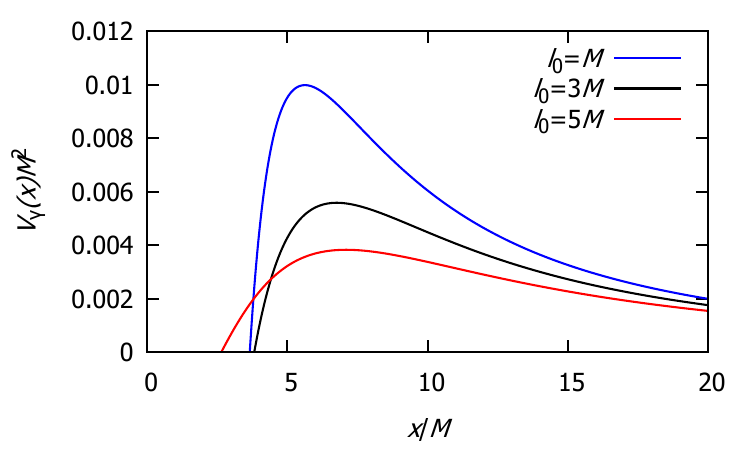}\hspace{-.1cm}
    \includegraphics[width=.5\linewidth]{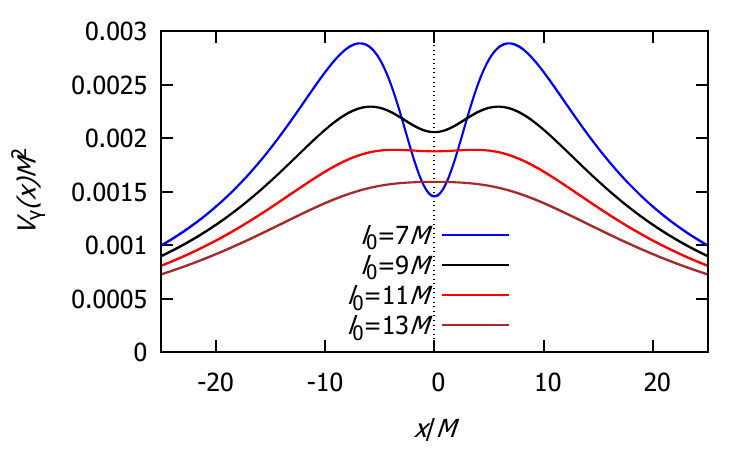}
    \caption{Effective potential for photons as a function of the radial coordinate for different values of $l_0$, for the case with  horizon (top panel) and the case without a horizon (bottom panel).}
    \label{fig:V}
\end{figure*}

\begin{figure}[htb]
    \centering
    \includegraphics[width=1\linewidth]{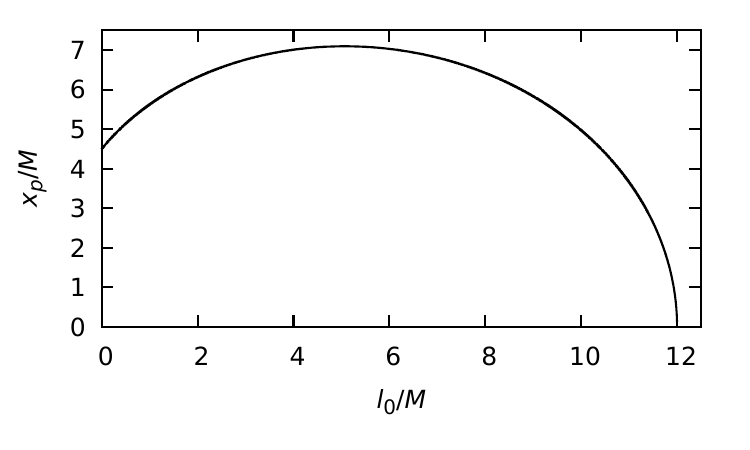}
    \caption{Behavior of the photon sphere radius as a function of $l_0$.}
    \label{fig:xpM}
\end{figure}

\begin{figure}[htb]
    \centering
    \includegraphics[width=1\linewidth]{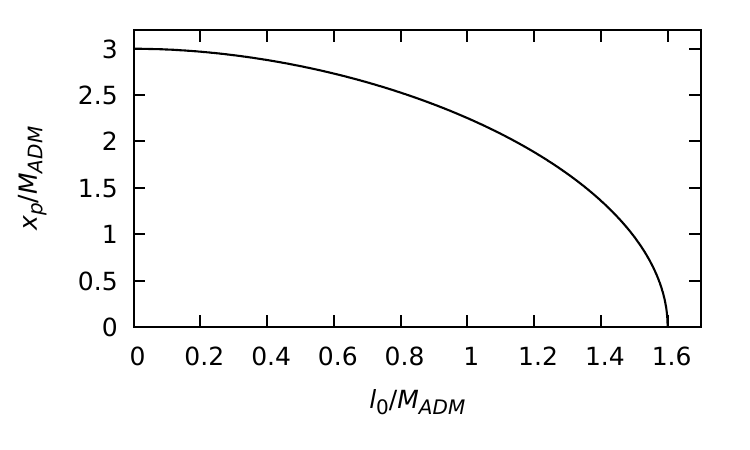}
    \caption{Behavior of the photon sphere radius as a function of $l_0$ when we normalize in terms of the ADM mass.}
    \label{fig:xpadm}
\end{figure}

It is important to note that the photon sphere, if it exists, is defined for $V_\gamma(x_p) = 1/b_c^2$, $V'_\gamma(x_p) = 0$ and $V''_\gamma(x_p) < 0$ (unstable orbit). Here, $b_c$ is the critical impact parameter, for which the photons undergo unstable circular orbits around the BB. In Fig. \ref{fig:V}, we have the plot of the effective potential for different values of the constant $l_0$. For both the horizon case (left panel) and the horizonless case (right panel), the plots reveal the existence of at least one unstable photon orbit in each scenario. However, for $l_0 >12M$,  the unstable photon orbit in the horizonless case becomes confined to the throat region only. This indicates that for $6M <l_0<12M$, unstable orbits still exist outside the throat. This result is better illustrated in Fig. \ref{fig:xpM}, where we observe the behavior of the photon sphere radius as the parameter $l_0$ is varied. In Table \ref{tabLR} we see some examples for the position of the unstable orbit for photons considering different values of $l_0$. Since the solution behaves similarly to Schwarzschild at infinity when the ADM mass is used in place of $M$ in the metric, we can analyze the behavior of the photon sphere radius when normalized by the ADM mass. This behavior is shown in Fig~\ref{fig:xpadm}.

\begin{widetext}
\centering
\begin{table*}[htb]
\begin{tabular}{lllllllll}
\cline{1-8}
\multicolumn{1}{c|}{}                  & \multicolumn{3}{c|}{Regular black hole}                                                                                          & \multicolumn{4}{c}{Wormhole}                                                                                                                                                    &                      \\ \cline{1-8}
\multicolumn{1}{c|}{}                  & \multicolumn{1}{c|}{$l_0=M$}           & \multicolumn{1}{c|}{$l_0 =3M$}         & \multicolumn{1}{c|}{$l_0=5M$}          & \multicolumn{1}{c|}{$l_0=7M$}          & \multicolumn{1}{c|}{$l_0=9M$}          & \multicolumn{1}{c|}{$l_0=11M$}         & \multicolumn{1}{c}{$l_0=13M$}                        &                      \\ \cline{1-8}
\multicolumn{1}{c|}{$x_p$}    & \multicolumn{1}{c|}{$\approx 5.625M$} & \multicolumn{1}{c|}{$\approx 6.757M$} & \multicolumn{1}{c|}{$\approx 7.094M$} & \multicolumn{1}{c|}{$\approx 6.808M$} & \multicolumn{1}{c|}{$\approx 5.830M$} & \multicolumn{1}{c|}{$\approx 3.658M$} & \multicolumn{1}{c}{$0$}                &                      \\ \cline{1-8}\cline{1-8}          
\end{tabular}
\caption{Position of the photon sphere for the black hole ($l_0 <6M$) and wormhole ($l_0 >6M$) cases, for different values of $l_0$.}
\end{table*}\label{tabLR}
\end{widetext}

It is important to highlight the role of the constant $l_0$
  in shaping the behavior of the effective potential for photons. For instance, in the horizonless case, the fact that values of $l_0>12M$ lead to the existence of a photon sphere confined exclusively to the throat suggests that quantum effects (associated with the constant $l_0$) might manifest at astrophysical scales. This is particularly significant, as the number and location of photon orbits have a direct impact on observable properties, such as the optical appearance and the shadow of these objects, if they exist at astrophysical scales.
  
In order to determine the actual trajectory of photons, we must derive an equation for the azimuthal angle $\phi$. This can be achieved by relating equation \eqref{xeq} to equation \eqref{killing}, which expresses the conservation of angular momentum. Proceeding in this way, we obtain
\begin{equation}
    \frac{d\phi}{dx} = \pm \frac{b}{\Sigma^2\sqrt{1 - \frac{b^2A}{\Sigma^2}}}.\label{dphidx}
\end{equation}

This equation is commonly used to determine the various types of trajectories that photons can follow around the object. Among other applications, it can be employed to analyze the optical appearance of the BB to an external observer when illuminated by a thin accretion disk.

\subsection{Shadow radius}
In this section, a comparison of the angular radius of the shadow of the supermassive black hole Sgr A*, as measured by the EHT, with the theoretical angular radius of the shadow computed from our black‑bounce solution, is performed. Despite the fact that the EHT does not measure the shadow radius directly, since it can observe just the 10\% highest peak of the luminosity, being unable to locate the angular aperture of the shadow, its value can be deduced by using independent data concerning the mass and the distance. Then, by requiring the theoretical radius to be consistent with the EHT observation within the experimental uncertainties, we derive constraints on the parameter $l_0$ that characterizes our solution, which, in principle, originates from the T-duality in string theory. In order to make this comparison, we implicitly assume that the observational appearance of the shadow, specifically, its angular radius as seen by an observer,is independent of the underlying metric or gravity theory. Consequently, enforcing the consistency between the observed and theoretical radii constrains only the parameter $l_0$, which until now has been treated as free. Therefore, our methodology enables observational limits, on astrophysical scales, to be placed on the constant $l_0$ arising from quantum corrections.
To apply this methodology, it is crucial to know the mass‑to‑distance ratio of Sgr A*. The mass $M$ and distance $D$ of Sgr A* have been extensively studied over the past decades through near-infrared (NIR) observations analysis, conducted with 8-10 $m$ class telescopes, tracking the orbits of so-called S-stars located in the immediate vicinity of Sgr A* \cite{EventHorizonTelescope:2022xqj}.

In this context, special attention has been given to the star S0-2, which has an apparent K-band magnitude $\sim$ 14, an orbital period $P \sim 14$ years, and a semi-major axis $a \sim 125 mas$. This star was primarily selected due to its short orbital period, which results in strong variations in its radial velocity and angular position, making it particularly sensitive to the values of $M$ and $D$.
S0-2 has been tracked with high precision by two sets of instruments: Keck and VLTI/GRAVITY (Very Large Telescope Interferometer).
The Keck telescope (a 10-meter telescope equipped with adaptive optics (AO) and the NIRC2 camera) achieves a positional uncertainty of about 0.5–1 $mas$, whereas VLTI/GRAVITY reaches a precision of 40–50 $\mu as$, thanks to the interferometric fringe-tracking technique.
For more details on the measurements, see Refs. \cite{EventHorizonTelescope:2022xqj,deMartino:2021daj,Jusufi:2021lei,Benisty:2021cmq,Borka:2021omc,Yuan:2022nmu,Do:2019txf,GRAVITY:2020gka,Vagnozzi:2022moj}. Table \ref{tab:mass_distance} presents the values of mass $M$ and distance $D$ obtained by both collaborations, along with their respective uncertainties.

\begin{table}[htb!]
\centering
\begin{tabular}{lcccl}
\hline
Survey & $M~(\times10^6 M_\odot)$ & $D$ (kpc) & Ref. \\
\hline
Keck & $3.951 \pm 0.047$ & $7.953 \pm 0.050 \pm 0.032$ & \cite{Do:2019txf} \\
VLTI & $4.297 \pm 0.012 \pm 0.040$ & $8.277 \pm 0.009 \pm 0.033$ & \cite{GRAVITY:2020gka} \\
\hline
\end{tabular}
\caption{Mass $M$ and distance $D$ to Sgr A* as measured by the Keck and VLTI surveys.}
\label{tab:mass_distance}
\end{table}

In addition to the mass-to-distance ratio, the so-called calibration factor must also be taken into account\cite{Vagnozzi:2022moj}. That is, since the EHT does not directly measure the shadow’s radius but rather the radius of the bright emission ring around the black hole, this factor corrects for the fact that what we observe is a slight offset from the true shadow radius, owing to emission processes, scattering, and photon trajectories. This factor is defined as \cite{Vagnozzi:2022moj,EventHorizonTelescope:2022xqj}:
\begin{equation}\label{delta}
    \delta \equiv \frac{r_{s}}{{r_{s}}^{schw}} - 1,
\end{equation}
where ${r_{s}}^{schw} = 3\sqrt{3}M$.

Thus, the calibration factor $\delta$ depends on the mass‑to‑distance ratio, yielding $\delta_{\rm Keck} = -0.04^{+0.09}_{-0.10}$ and $\delta_{\rm VLTI} = -0.08^{+0.09}_{-0.09}$ \cite{Vagnozzi:2022moj,EventHorizonTelescope:2022xqj}. Since these two estimates are independent, we take their average, giving
\begin{equation}
     \delta \simeq -0.060 \pm 0.065\,, 
\end{equation}
which finally leads us to the following confidence intervals \cite{Vagnozzi:2022moj}:
\begin{equation}
     -0.125 \lesssim \delta \lesssim 0.005 \; (1\sigma),
  \quad
  -0.19 \lesssim \delta \lesssim 0.07 \; (2\sigma).
\end{equation}

By examining the values obtained from the Keck and VLTI data, one can conclude that the calibration factor $\delta$ is slightly negative, indicating that the observed shadow radius tends to be smaller than the true value, within an approximately 68\% confidence interval \cite{Vagnozzi:2022moj}. If we invert relation \eqref{delta}, we obtain:
\begin{equation}
    \frac{r_{s}}{M} = 3\sqrt{3}\,\,(1 + \delta),
\end{equation}
which implies the following intervals for the observed shadow radius \cite{Vagnozzi:2022moj}:
\begin{equation}
    4.55 \lesssim \frac{r_{s}}{M} \lesssim 5.22 \quad (1\sigma), 
    \quad
    4.21 \lesssim \frac{r_{s}}{M} \lesssim 5.56 \quad (2\sigma).
\end{equation}

Next, we will use these results to constrain the parameter $l_0$ of our solution.

\begin{figure}[htb!]
    \centering
    \includegraphics[width=1\linewidth]{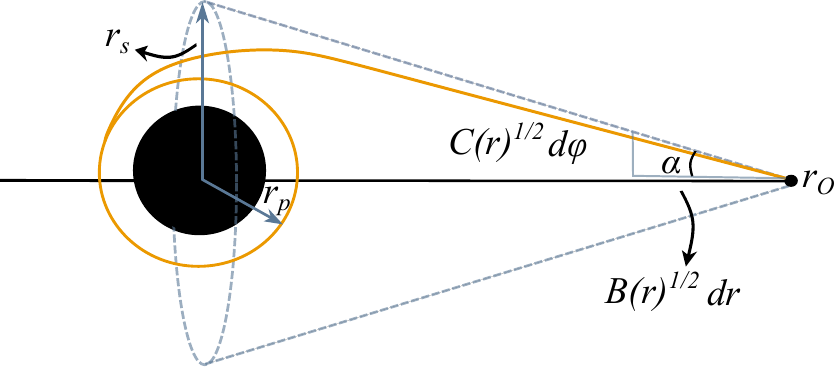}
    \caption{Representation of a light ray emitted from the point $r_O$ and entering a circular orbit around the black hole.}
    \label{fig:shapeshadow}
\end{figure}

To compute the theoretical shadow radius, we follow the formulation presented in \cite{Perlick:2021aok}. Accordingly, we consider a line element given by:
\begin{equation}
    ds^2=-A(r)dt^2+B(r)dr^2+C(r)d\Omega_2^2.
\end{equation}
The coordinates $x$ and $r$ are related by $r^2=\Sigma(x)^2$. Based on Fig. \ref{fig:shapeshadow}, we can relate the angle under which a photon ending in an unstable orbit is emitted to the metric components as:
\begin{equation}
\begin{split}
    \cot \alpha = \frac{\sqrt{g_{rr}}}{\sqrt{g_{\varphi\varphi}}}\left. \frac{dr}{d\varphi}\right|_{r=r_O}=\frac{\sqrt{B(r)}}{\sqrt{C(r)}}\left. \frac{dr}{d\varphi}\right|_{r=r_O}\\=\frac{\sqrt{B(r)}}{r}\left. \frac{dr}{d\varphi}\right|_{r=r_O},\label{cotalpha}
    \end{split}
\end{equation}
where $r_O$ is the point at which we place our observer. In fact, it can be interpreted either as the point from which the photons are emitted toward the black hole or as the location where the observer receives the photons that are absorbed by the black hole.

By performing the appropriate coordinate transformation in equation \eqref{dphidx} together with the relation proposed in equation \eqref{cotalpha}, it is possible to show that
\begin{equation}
    \sin^2\alpha = \frac{b^2A(r_O)}{r_O^2}.
\end{equation}
For small angles, we can use the approximation $\sin\alpha \approx \alpha$, and we get
\begin{equation}
    \alpha^2=\frac{b^2A(r_O)}{r_O^2}.
\end{equation}

In the small angle limit, we can also relate the angle, the observer's distance, and the shadow radius through the relation
$\alpha \approx r_{s}/r_O$.

For the case of the circular photon orbit, we have $b = b_c$, which, as previously mentioned from equations \eqref{radial_velocity} and \eqref{V_gen}, is given by $b_c^2 = 1/V(x_p) = \Sigma(x_p)^2/A(x_p)$. However, one must be careful to consider the appropriate coordinate system. With the necessary adjustments, the apparent shadow radius is written as
\begin{equation}
    r_S=r_p\sqrt{\frac{A(r_O)}{A(r_p)}}.
\end{equation}

\begin{figure*}[htb]
    \centering
    \includegraphics[width=.5\linewidth]{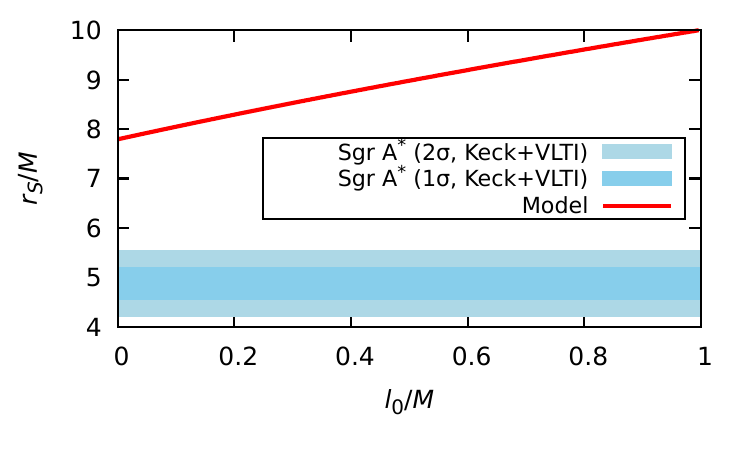}\hspace{-.1cm}
    \includegraphics[width=.5\linewidth]{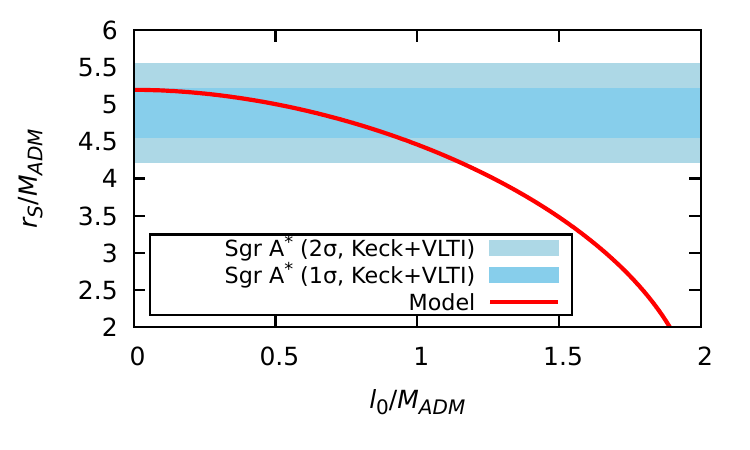}
    \caption{Apparent shadow radius of the black hole for the cases where $r_s/M$ (top panel) and $r_s/M_{ADM}$ (bottom panel) are considered as a function of the parameter $l_0$.}
    \label{fig:shadows_radius}
\end{figure*}

In Fig. \ref{fig:shadows_radius}, we compare the shadow mass-to-distance ratio for our model, considering both the mass $M$ and the ADM mass. We see that, when considering the ratio $r_s/M$, there is no value of $l_0/M$ for which this ratio falls within the acceptable ranges for $1\sigma$ or $2\sigma$. In the second case, when we consider the ratio $r_s/M_{\text{ADM}}$, we observe that there are values of $l_0/M_{\text{ADM}}$ for which our model lies within the error bounds, both for $1\sigma$ and $2\sigma$. In fact, for $l_0 \lesssim 1.15\, M_{\text{ADM}}$, the shadow radius predicted by our model falls within the $2\sigma$ confidence interval.

\subsection{Optical appearance}

\begin{figure}[htb]
    \centering
    \includegraphics[width=1.0\linewidth]{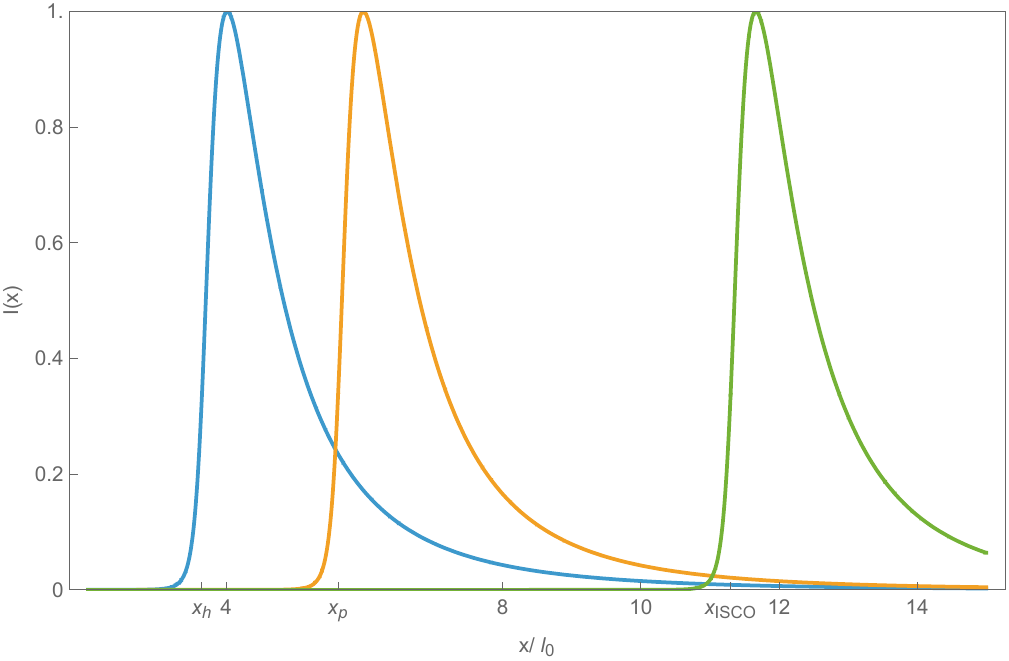}
    \caption{Intensity profiles according to the distribution \eqref{IntensitySU}. The luminosity peaks are normalized to the unity. Here, we are assuming $\gamma=-2$, $\sigma=1/4$ and the peak located at event horizon, $\mu=3.65$ (blue line), at the photon sphere, $\mu=5.62$ (orange line), and at the ISCO $\mu=11.3$ (blue line). The latter are calculated for a regular black hole with $l_0/M=1$.}
    \label{fig:Intensity_Profiles}
\end{figure}

\begin{figure*}[htb]
    \centering
    \includegraphics[width=0.4\linewidth]{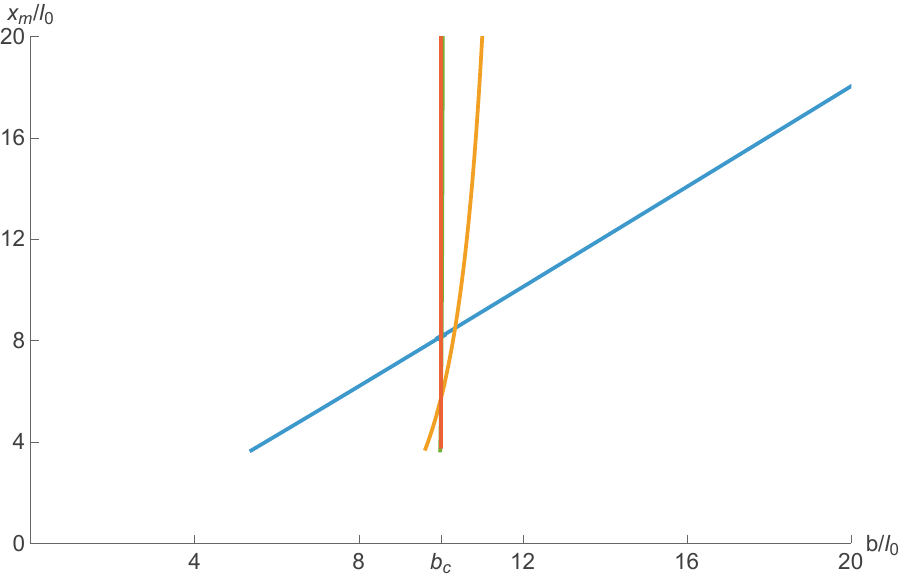}
        \includegraphics[width=0.4\linewidth]{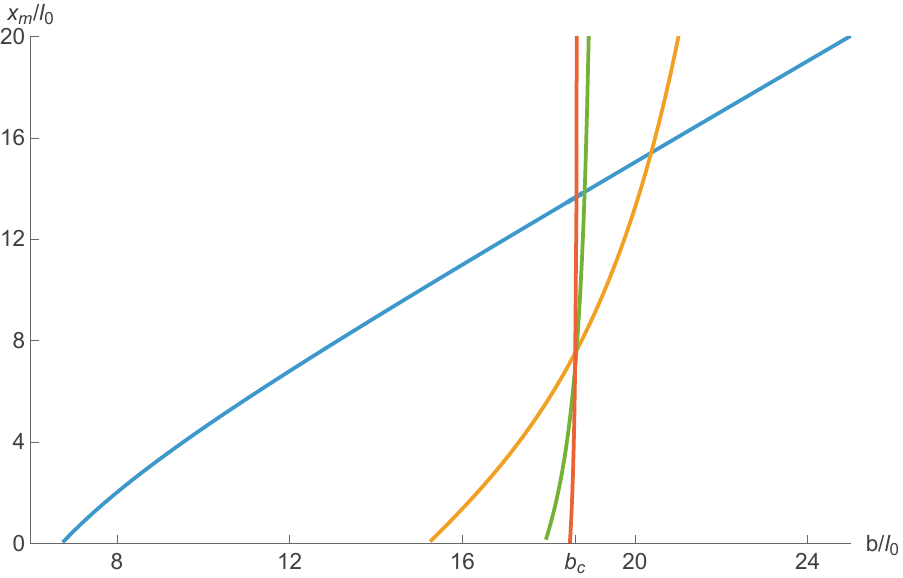}
    \caption{Transfer functions for the case of a regular black hole $l_0/M=1$), a wormhole ($l_0/M=7$). The lines correspond to photons going directly to the observer (blue), those going one half-turn (orange), two half-turns (green) and three half-turns (red).}
    \label{fig:Transfer_Func}
\end{figure*}

\begin{figure*}[htb]
    \centering
    \includegraphics[width=0.4\linewidth]{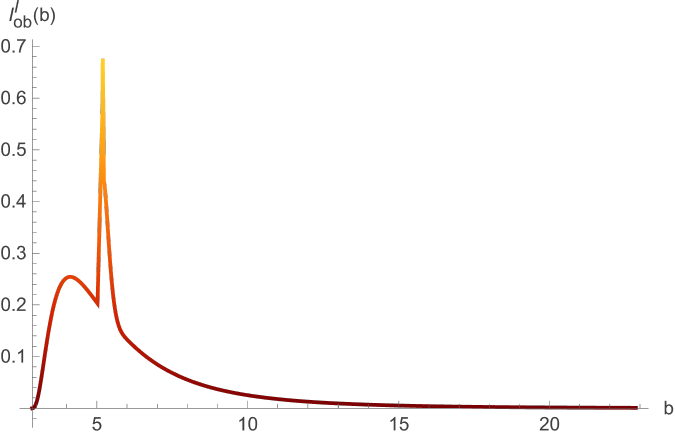}
        \includegraphics[width=0.4\linewidth]{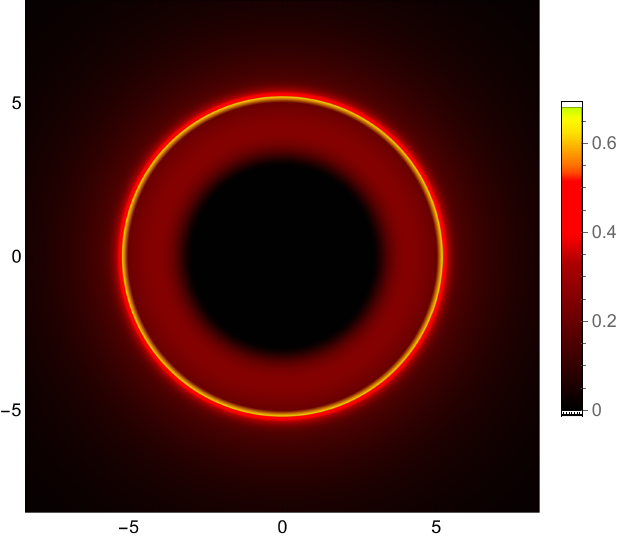}
\includegraphics[width=0.4\linewidth]{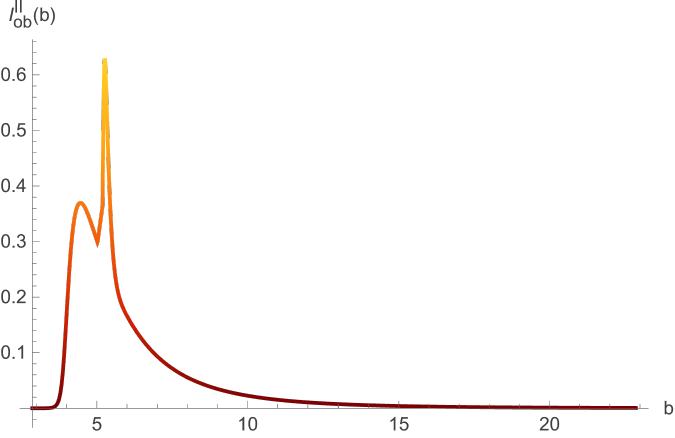}
        \includegraphics[width=0.4\linewidth]{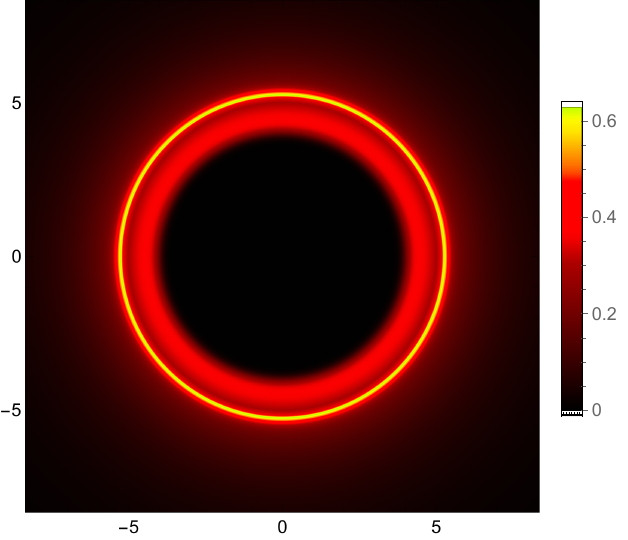}
        \includegraphics[width=0.4\linewidth]{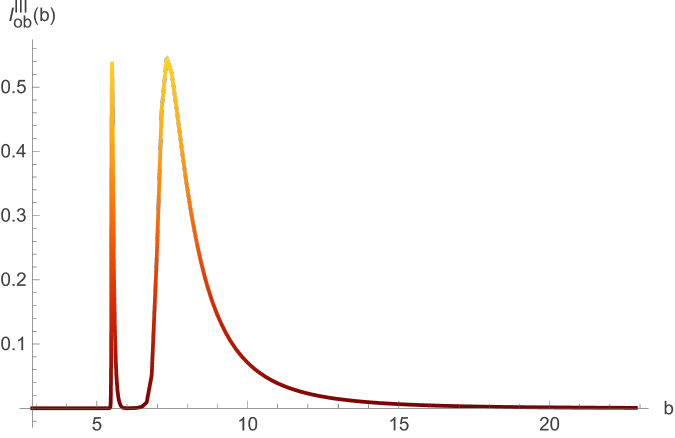}
           \includegraphics[width=0.4\linewidth]{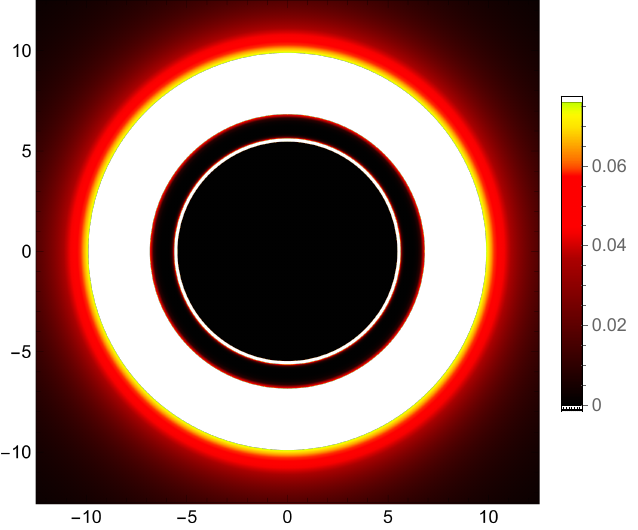}
    \caption{Observed luminosities (left figures) and optical appearance (right panels) for Schwarzschild black hole. From top to the bottom, the peak of the luminosity is located at the event horizon ($x_{H}/M=2$), at the photon sphere ($x_{p}/M=3$)  and at the ISCO ($x_{ISCO}/M=6$).}
    \label{fig:Optical_Appe_SCH}
\end{figure*}

\begin{figure*}[htb]
    \centering
    \includegraphics[width=0.4\linewidth]{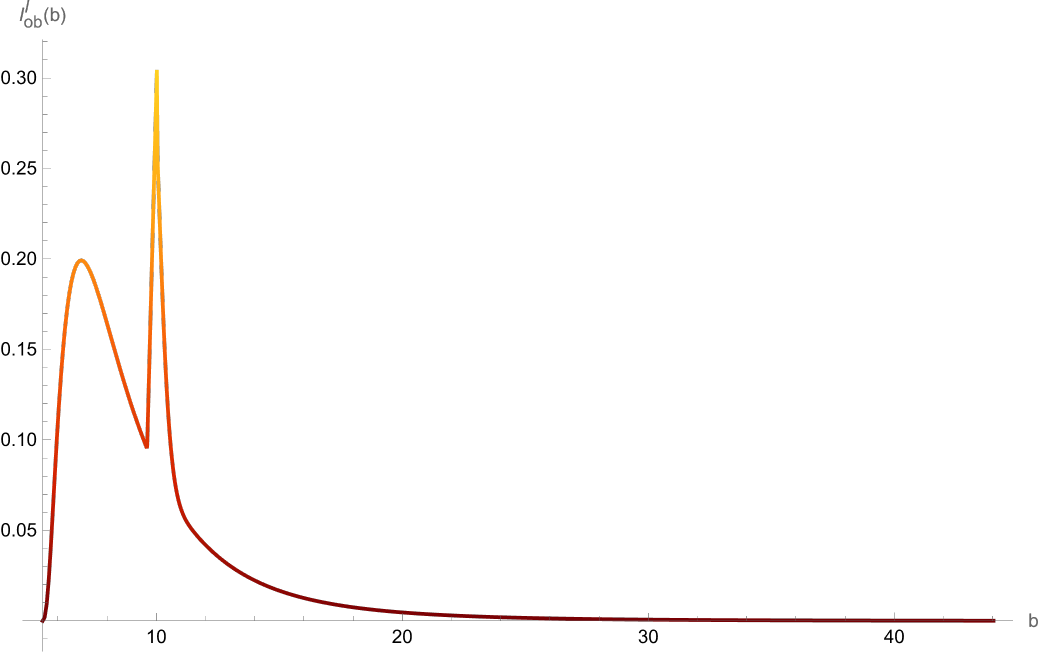}
        \includegraphics[width=0.4\linewidth]{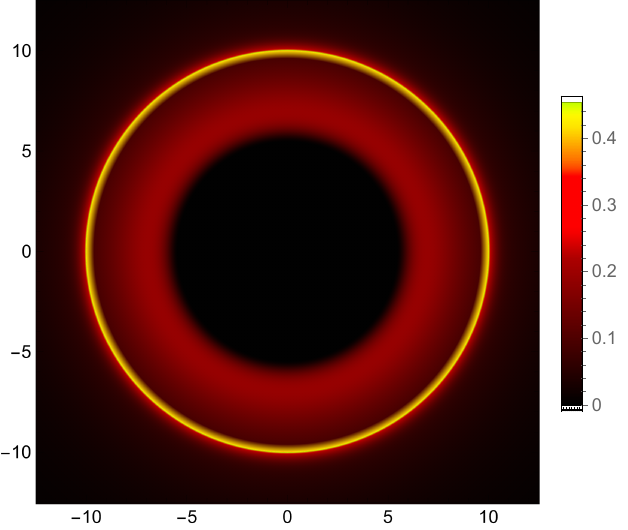}
\includegraphics[width=0.4\linewidth]{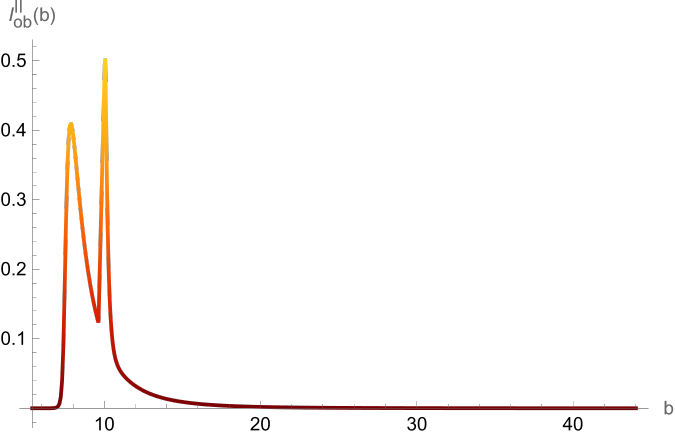}
        \includegraphics[width=0.4\linewidth]{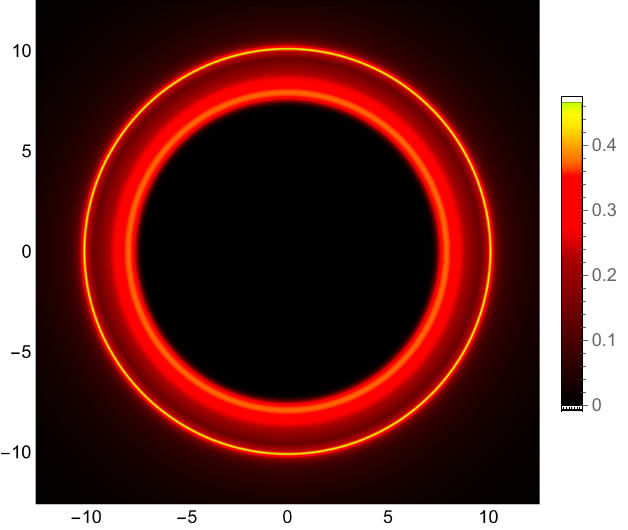}
        \includegraphics[width=0.4\linewidth]{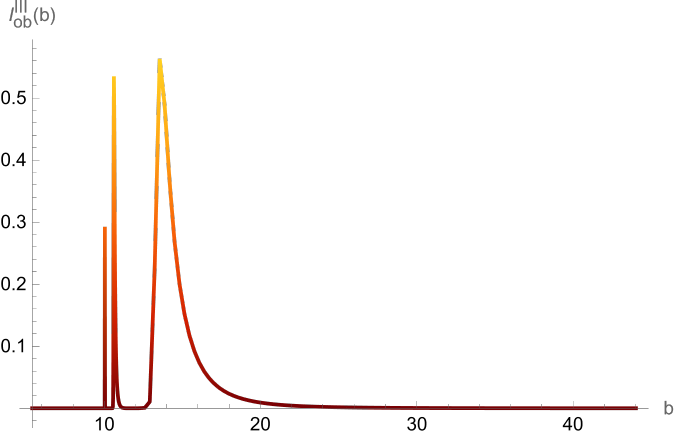}
           \includegraphics[width=0.4\linewidth]{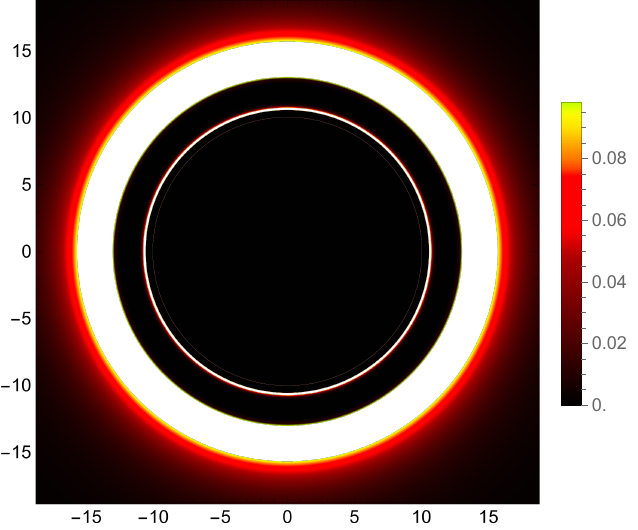}
    \caption{Observed luminosities (left figures) and optical appearance (right panels) for a regular black hole described by the spacetime metric \eqref{Ax} for $l_0/M=1$. From top to the bottom, the peak of the luminosity is located at the event horizon ($x_H/l_0\approx3.65$), at the photon sphere ($x_{p}/l_0\approx5.62$) and at the ISCO ($x_{ISCO}/M\approx11.3$).}
    \label{fig:Optical_Appe_RBH}
\end{figure*}

\begin{figure*}[htb]
    \centering
    \includegraphics[width=0.4\linewidth]{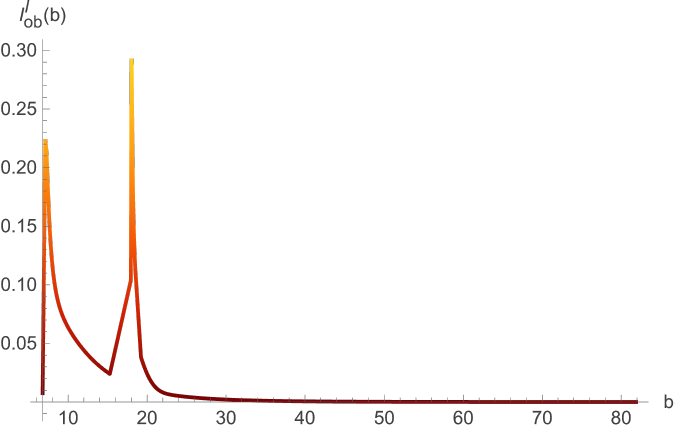}
        \includegraphics[width=0.4\linewidth]{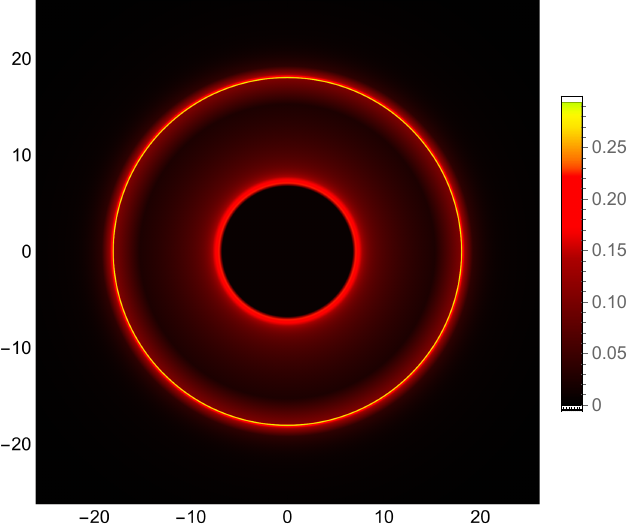}
\includegraphics[width=0.4\linewidth]{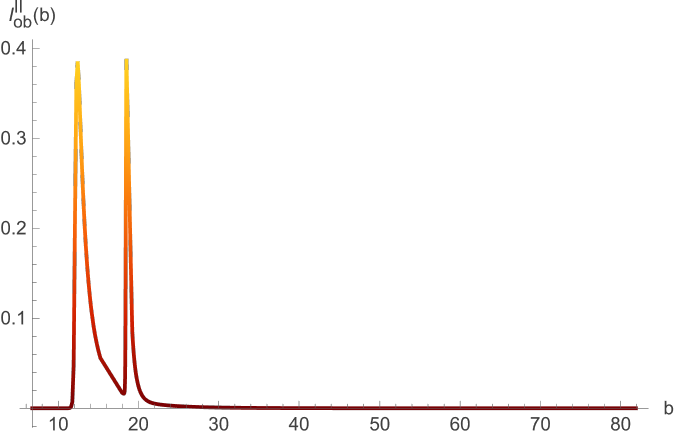}
        \includegraphics[width=0.4\linewidth]{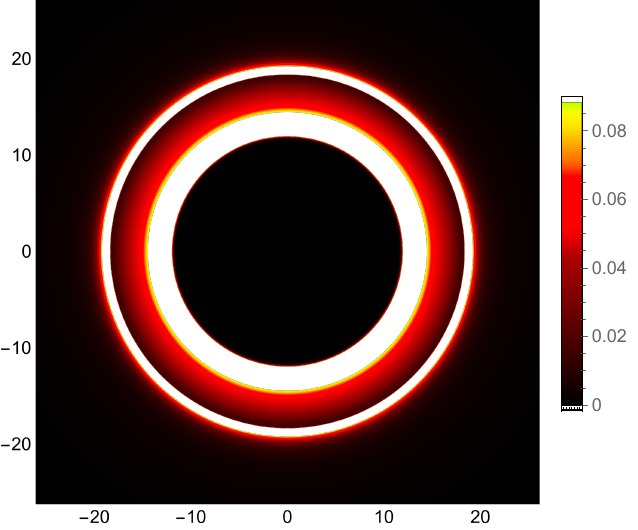}
        \includegraphics[width=0.4\linewidth]{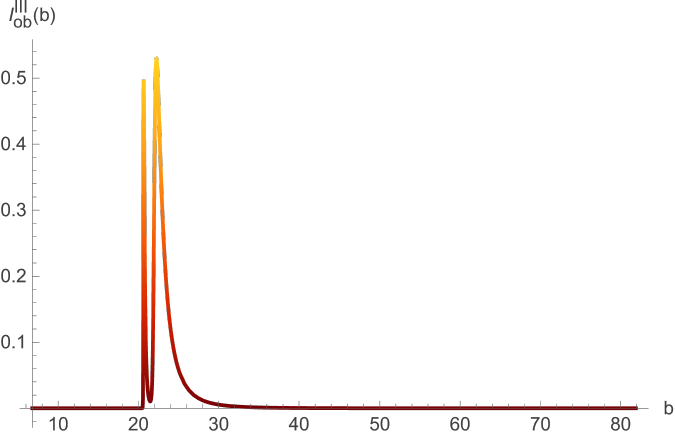}
           \includegraphics[width=0.4\linewidth]{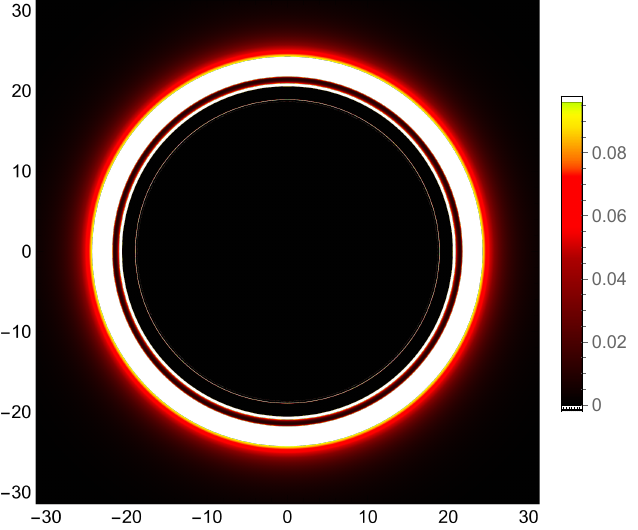}
    \caption{Observed luminosities (left figures) and optical appearance (right panels) for a wormhole described by the spacetime metric \eqref{Ax} with $l_0/M=7$. From top to the bottom, the peak of the luminosity is located just over the throat ($x=0$), at the photon sphere ($x_{p}/l_0\approx5.62$) and at the ISCO ($x_{ISCO}/l_0\approx16.9$).}
    \label{fig:Optical_Appe_WH}
\end{figure*}

Let us now simulate the optical appearance of such ultra-compact objects as seen by a distant observer. To do so, the emission for the accretion disk, which will be assumed to be optically and geometrically thin, is described by a particular luminosity profile. In addition, we locate the disk at the equatorial plane and consider that it emits from a certain inner edge outward. The intensity profile is considered to be isotropic and depends only on the radial coordinate. Furthermore, the effects associated with the emission and absorption coefficients of the relativistic radiative transfer equation are assumed to be negligible. Then, a Standard Unbound distribution is considered, which gives the following luminosity profile \cite{Paugnat:2022qzy},
\begin{equation}
    I^{em}_{SU}(x,\mu,\sigma,\gamma)=\frac{e^{-\frac{1}{2}\left[\gamma+\text{arc
    sinh}\left(\frac{x-\mu}{\sigma}\right)\right]^2}}{\sqrt{(x-\mu)^2+\sigma^2}},
        \label{IntensitySU}
\end{equation}
where the parameters $\mu$, $\sigma$, and $\gamma$ define the position of the emission peak, the scale, and the asymmetry of the central region, respectively. Here, the values of these free parameters are chosen such that the emission peak corresponds to the event horizon (if any), the photon sphere, and the ISCO. In Fig.~\ref{fig:Intensity_Profiles}, such profiles for  $I^{em}_{SU}$ as a function of $x$ are depicted. In order to reconstruct the images as the average of collected photons that reach the observer's screen, the intensity \eqref{IntensitySU} is affected by two effects. Firstly, the gravitational redshift, which is determined by the position $x$ in the accretion disk at which the photon is emitted, since its frequency varies as $\nu' = \left(\frac{A(x)}{A(x_{ob})}\right)^{1/2}\nu$ and consequently, the observed intensity turns out
\begin{equation}
I^{ob}_{\nu'} = \left(\frac{A(x)}{A(x_{ob})}\right)^{3/2} I^{em}_{\nu}\ .
\end{equation}
Secondly, the strong deflection of photon trajectories, exacerbated by the presence of a circular orbit for photons, may make photons turn around the central object and cross the accretion disk more than once before running away towards the observer. This effect contributes to the brightness of the optical appearance. Hence, one has to sum such additional intensities to obtain the actual observed luminosity,
\begin{equation}
    I^{obs}(b)=\frac{1}{A(x_{ob})^2}\sum_n [A^2I]\vert_{x=x_n(b)}\ ,
\end{equation}
where $x_n(b)$ is the transfer function that relates the position at which the photon crosses the equatorial plane and its impact parameter. Figure \ref{fig:Transfer_Func} shows the transfer functions for the cases where spacetime \eqref{Ax} describes a regular black hole and a wormhole, respectively. Then, the optical appearance is simulated by considering two particular cases for the BB metric \eqref{Ax}: a regular black hole ($l_0/M=1$) and a wormhole ($l_0/M=7$). For each case, we consider the three luminosity profiles depicted in Fig.~\ref{fig:Intensity_Profiles}, except the one whose emission peak is located at the event horizon for the wormhole, which we locate over the throat. Both cases are depicted in Figs.~\ref{fig:Optical_Appe_RBH} and \ref{fig:Optical_Appe_WH}. For comparison, we have also simulated the optical appearance for the Schwazrschild black hole, depicted in Fig.~\ref{fig:Optical_Appe_SCH}, and the corresponding image for a Bardeen-like wormhole whose metric is given by \eqref{metric} but the radius of the $2-$spheres is $r^2=x^2+l_0^2$, which might be seen as a competitor to the wormhole described by \eqref{Ax}. \\

\begin{figure*}[htb]
    \centering
    \includegraphics[width=0.4\linewidth]{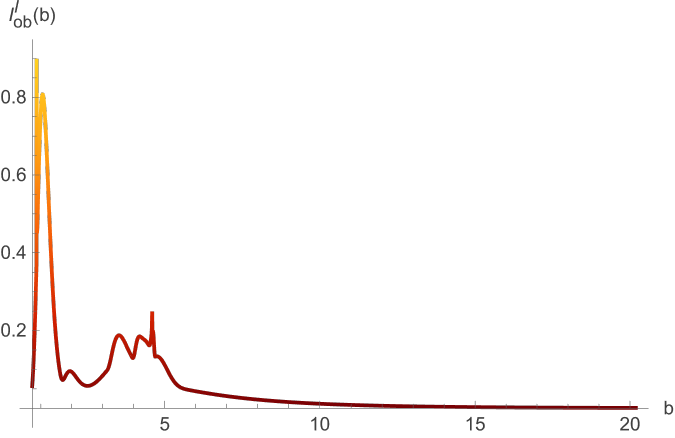}
     \includegraphics[width=0.4\linewidth]{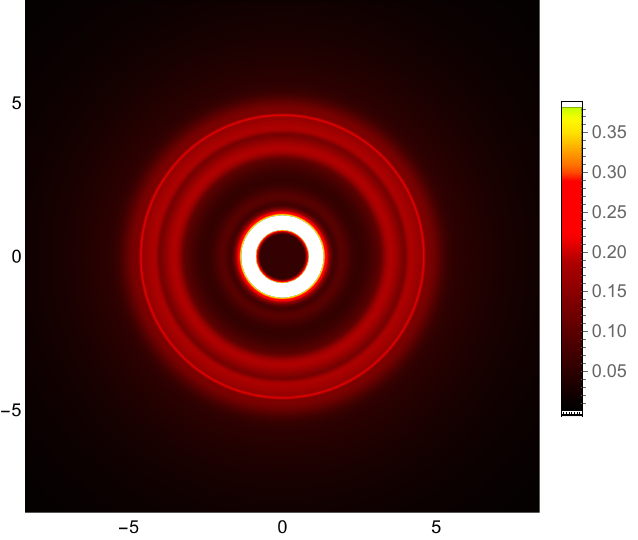}
\includegraphics[width=0.4\linewidth]{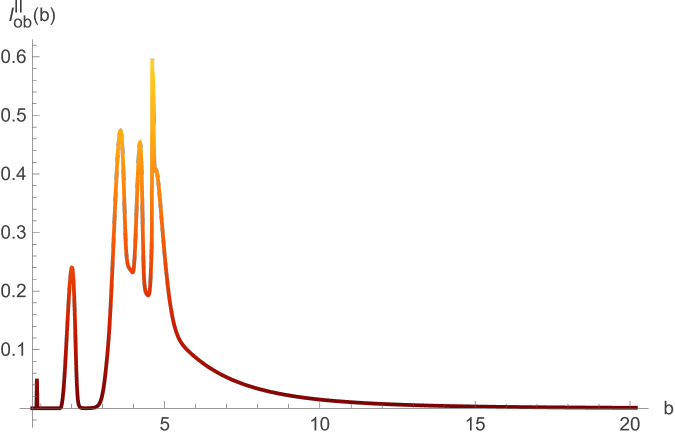}
        \includegraphics[width=0.4\linewidth]{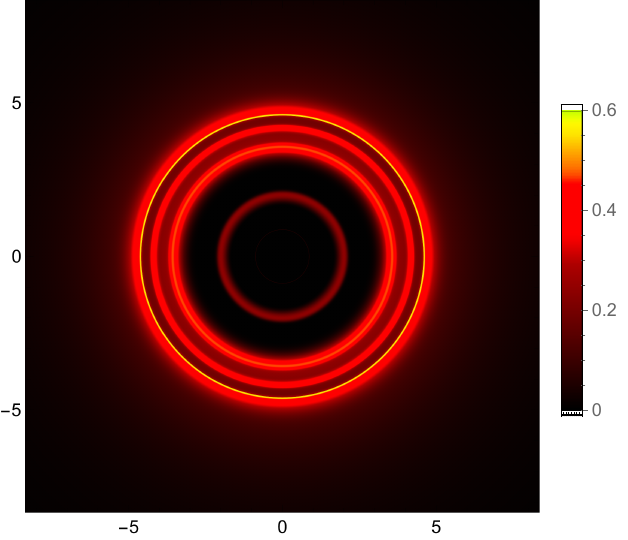}
        \includegraphics[width=0.4\linewidth]{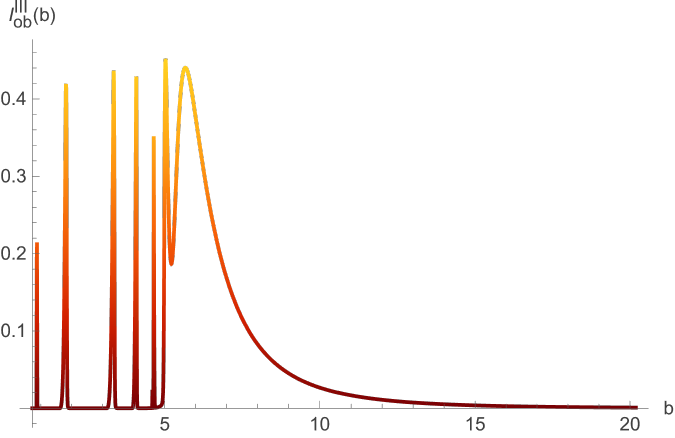}
           \includegraphics[width=0.4\linewidth]{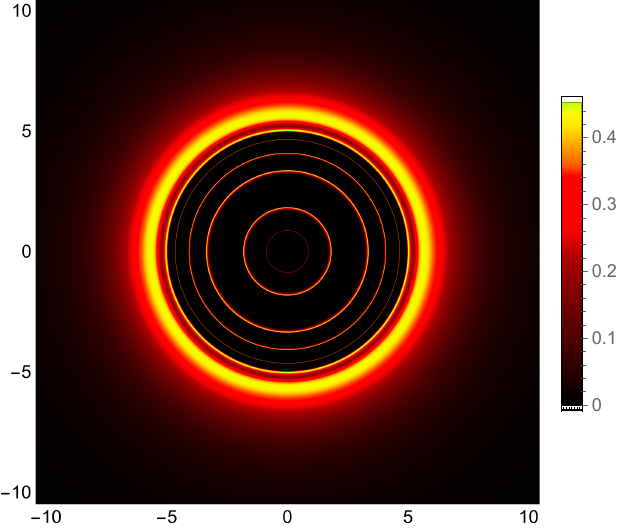}
    \caption{Observed luminosities (left figures) and optical appearance (right panels) for the wormhole \eqref{metric} with $l_0/M=6/7$. From top to the bottom, the peak of the luminosity is located at the throat ($x_{T}/M=0$), at the photon sphere ($x_{p}/M=2.15$)  and at the ISCO ($x_{ISCO}/M=4.29$).}
    \label{fig:Optical_WHLobo}
\end{figure*}

Hence, when comparing black holes described by the classical Schwazrschild metric, Fig.~\ref{fig:Optical_Appe_SCH}, and the regular black hole given by the novel spacetime metric \eqref{Ax} and its shadow depicted in Fig.~\ref{fig:Optical_Appe_RBH}, one can infer some differences on the width of the light rings. However, such a difference might be indistinguishable from other effects coming from the accretion disk, such as, for instance, different luminosity profiles. Nevertheless, there are clear differences between the case of a regular black hole, Fig.~\ref{fig:Optical_Appe_RBH}, and the wormhole described by the same metric,  Fig.~\ref{fig:Optical_Appe_WH}. Then, by comparing the wormholes described by different metrics, Figs.~\ref{fig:Optical_Appe_WH} and \ref{fig:Optical_WHLobo}, one can see that there are clear differences in the optical appearance. This is a consequence of the presence of multiple photon circular orbits for the case of the Bardeen-like wormhole, which induce multiple photon rings in the image, whereas the metric \eqref{Ax} just owns a unique photon sphere, as pointed out in the previous sections, so the photon rings structure is much simpler than the Bardeen-like wormhole but distinguishable from the regular black hole case.

\section{Thermodynamics}\label{S:HCS}
\subsection{Hawking temperature}
We can now study the thermodynamics of the regular black hole case, where a horizon exists, that is, the parameter $l_0$ lies within the interval $l_0 \in (0, 6M]$. We begin by computing the surface gravity $\kappa_{\mathcal{H}} $ at the event horizon $\mathcal{H}$, that is, on the null hypersurface that defines the Killing horizon of the manifold, where in this case the Killing vector normal to the hypersurface $\mathcal{H}$ is simply given by the timelike translation Killing vector $\bm{\xi} = \partial_t$,where the norm is simply given by
\begin{equation}
   \left|\left| \bm{\xi} \right|\right|^2 = \xi^\gamma \xi_\gamma = g_{\alpha\beta}\xi^\alpha\xi^\beta = -A(x).
\end{equation}

Given the expression for the surface gravity,
\begin{equation}
    \nabla_\mu\left(\xi^\alpha\, \xi_\alpha\right) = -2\kappa_{\mathcal{H}}\, \xi_\mu,
\end{equation}
we finally obtain the expression
\begin{equation}
    \kappa_{\mathcal{H}}= \frac{1}{2}A'|_{x=x_H} = \frac{1}{2x_H^3}\left[x_H^2 + 2l_0^2\left(\frac{l_0}{\sqrt{x_H^2 + l_0^2}} - 1\right)\right],
\end{equation}
recalling that $x_H$ is written as \eqref{xh}. In the geometrodynamics units where $ G=\hbar=k_B=c=1$, the Hawking temperature is defined as $T_H = \kappa_{\mathcal{H}}/2\pi$, witch gives
\begin{equation}
    T_H(x_H) = \frac{1}{4\pi x_H^3}\left[x_H^2 + 2l_0^2\left(\frac{l_0}{\sqrt{x_H^2 + l_0^2}} - 1\right)\right].
\end{equation}
From this expression, we can analyze some thermodynamic characteristics of our solution. For example, for values of $x_H \to 0$ (the final stages of evaporation), we observe the following behavior of the temperature:
\begin{equation}
    T_H(x_H \to 0)\approx \frac{3x_H}{16\pi l_0^2} + O(x_H^3),
\end{equation}
which therefore allows us to conclude that for $x_H = 0$, we will have $T_H = 0$, meaning that in this solution the complete evaporation of the object is possible, unlike what happens in Schwarzschild black hole. By the other hand, for the asymptotic case where $x_H \to \infty$, we have the following behavior:
\begin{equation}
    T_H(x_H \to \infty) \approx \frac{1}{4\pi\, x_H} + O(1/x_H^3),
\end{equation}
which is precisely the Hawking temperature for the Schwarzschild case,which indicates that in the early stages of evaporation, our solution behaves thermodynamically in a manner analogous to Schwarzschild, with the quantum correction effects arising from string T-duality becoming relevant only in the final stages of evaporation.

\begin{figure}[htb]
    \centering
    \includegraphics[width=1\linewidth]{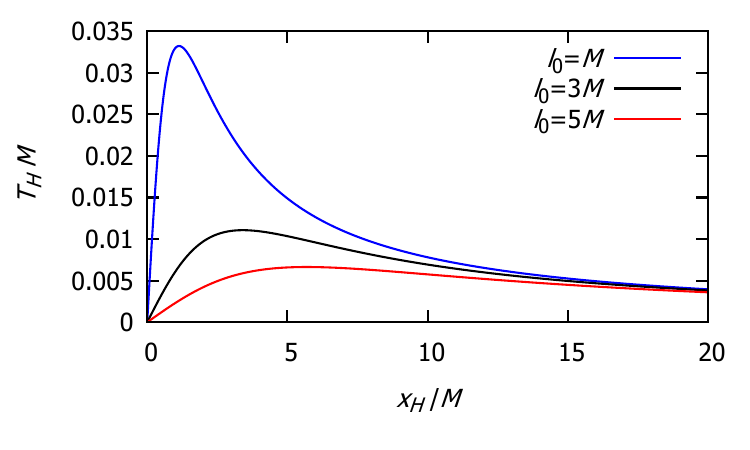}
    \caption{Hawking temperature as a function of the horizon radius for different values of $l_0$.}
    \label{fig:T}
\end{figure}

The complete behavior of the Hawking temperature for this solution can be graphically analyzed by observing Fig. \ref{fig:T}. In the plot, we can qualitatively observe the two behaviors discussed earlier. That is, near $x_H = 0$, the temperature behaves approximately linearly as a function of $x_H$, with a slope that depends on the constant $l_0$; for large values of $x_H$, the temperature behaves inversely proportional to $x_H$, analogous to the Schwarzschild case. Moreover, it can be seen that the temperature remains finite throughout the entire evaporation process and reaches a maximum at a certain critical value of the horizon radius. This will have implications for the behavior of the heat capacity, as we shall see in the following.

\subsection{Heat capacity}

In black hole thermodynamics, the heat capacity provides information about the thermodynamic stability of the system, such that stability is achieved when $C \geq 0$ \cite{stanley,davies1977,Rodrigues:2022qdp}. The heat capacity is then defined as
\begin{equation}
    C = \frac{dQ}{dT_H}= T_H \frac{dS}{dT_H} = T_H\left(\frac{dS}{dx_H}\right)\left(\frac{dT_H}{dx_H}\right)^{-1},
\end{equation}
where $dQ$ is the heat that is emitted or absorbed by the BB, and $S$  is the entropy that satisfies the area law $S = A_{\mathcal{H}}/4 = \pi (x_H^2 + l_0^2)$. We can explicitly write the heat capacity in terms of the horizon radius, obtaining
\begin{align}
C =\; & 2\pi x_H^2 (l_0^2 + x_H^2) \times \notag \\
& \frac{3 l_0^4 + 6 l_0^2 x_H^2 - x_H^4 + 6 l_0^3 \sqrt{l_0^2 + x_H^2}}
       {9 l_0^6 + 3 l_0^4 x_H^2 - 9 l_0^2 x_H^4 + x_H^6},
\end{align}
or, alternatively, write it in terms of the entropy, where we obtain
\begin{equation}
    C =  \frac{2S \left(2 l_0 \sqrt{\pi} + \sqrt{S} \right) \left(S -l_0^2 \pi\right)}{2 l_0^3 \pi^{3/2} + 4 l_0^2 \pi \sqrt{S} - 2 l_0 \sqrt{\pi} S - S^{3/2}}.
\end{equation}

\begin{figure}[htb]
    \centering
    \includegraphics[width=1\linewidth]{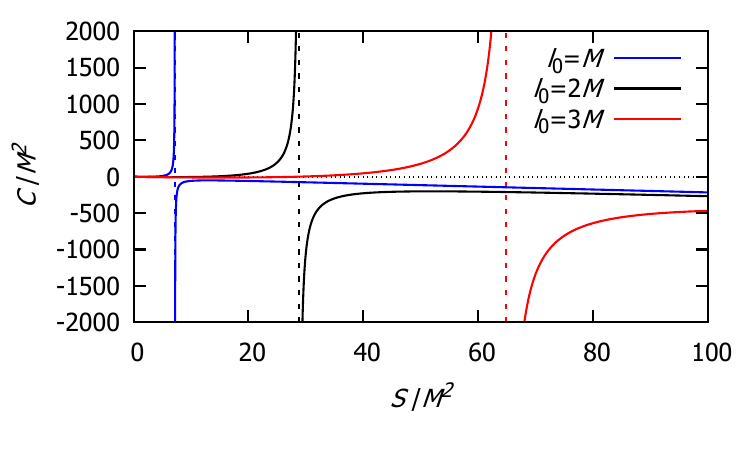}
    \caption{Heat capacity as a function of entropy for different values of $l_0$.}
    \label{fig:C}
\end{figure}

Upon examination of the graph of the heat capacity as a function of entropy in Fig. \ref{fig:C}, we can see that there is a divergence in the heat capacity for each value of $l_0$. These critical points, where the heat capacity diverges, are marked by $dT_H/dS=0$, which indicates a second-order phase transition. The location of this transition, in terms of entropy, increases as the parameter $l_0$ increases.

Thus, the evaporation process of the BB can be summarized as follows: In the beginning, when the event horizon is considerably large, the Hawking temperature is low and the heat capacity is negative, indicating that, as it radiates, the BB increases its temperature and accelerates its mass loss, exhibiting an unstable behavior identical to that of the Schwarzschild black hole. Eventually, after crossing the critical point where the phase transition occurs, the heat capacity becomes positive, entering a window of thermal stability that slows the evaporation. In this phase, the emission of energy leads to a decrease in temperature, which makes the evaporation process self-regulating, until the BB completely evaporates without leaving behind any remnant.

It is also important to note that, as the value of the parameter $l_0$ increases, the location of the divergence in entropy in the heat capacity graph changes to higher values. This implies that the window of thermal stability becomes broader, indicating that BBs with larger $l_0$ remain stable over a longer portion of their thermodynamic evolution. In this way, the parameter $l_0$, which originates from quantum corrections, acts as a kind of regulator of BB thermodynamics, playing a stabilizing role during the evaporation process.

\section{Energy Conditions}\label{S:EnerCond}
In order to assess how exotic these sources are, we must analyze the energy conditions. The conditions most commonly considered are the null energy condition (NEC), the weak energy condition (WEC), the strong energy condition (SEC), and the dominant energy condition (DEC). These conditions are given by:
\begin{eqnarray}
&&NEC_{1,2}=WEC_{1,2}=SEC_{1,2}
\Longleftrightarrow \rho+p_{r,t}\geq 0,\label{Econd1} \\
&&SEC_3 \Longleftrightarrow\rho+p_r+2p_t\geq 0,\label{Econd2}\\
&&DEC_{1,2} \Longleftrightarrow \rho-p_{r,t}\geq 0, \quad \mbox{and} \quad \rho+p_{r,t}\geq 0, \label{Econd3}\\
&&DEC_3=WEC_3 \Longleftrightarrow\rho\geq 0.\label{Econd4}
\end{eqnarray}
Each condition involves a set of inequalities, and some of these inequalities appear in more than one condition. Inequality $\rho + p_r$, for example, appears in NEC, WEC, SEC, and DEC. Therefore, if this inequality is not satisfied, all energy conditions are violated. Even when the energy conditions are violated, some of the individual inequalities may still be satisfied. For the DEC, we will consider only the conditions $DEC_{1,2} \Longrightarrow \rho - p_{r,t} \geq 0$, since the other inequality is already included in the NEC.

To properly analyze the energy conditions, we shall once again examine the components of the stress-energy tensor. However, care must be taken in identifying the regions where $t$ behaves as a timelike or spacelike coordinate, so that the physical interpretation of the quantities $\rho$, $p_r$, and $p_t$ remains consistent with the metric signature. The fluid quantities are related to the components of the stress-energy tensor as
\begin{eqnarray}
T^{\mu}{}_{\nu}={\rm diag}\left[-\rho,p_r,p_t,p_t\right]\, \quad \mbox{if} \quad A(x)>0,\label{EMT}\\
T^{\mu}{}_{\nu}={\rm diag}\left[p_r,-\rho,p_t,p_t\right]\, \quad \mbox{if} \quad A(x)<0.\label{EMTOut}
\end{eqnarray}

Inequality $NEC_1$ is written as
\begin{equation}
\begin{split}
NEC_{1}
\Longleftrightarrow -\frac{2A\Sigma''}{\Sigma}, \quad \mbox{if} \quad A(x)>0,\\
NEC_{1}
\Longleftrightarrow \frac{2A\Sigma''}{\Sigma}, \quad \mbox{if} \quad A(x)<0.
\end{split}
\end{equation}
For our model, we always have $\Sigma > 0$ and $\Sigma'' > 0$, such that the inequality $NEC_1$ is always violated.

In the regions where $A(x) > 0$, the energy density is given by \eqref{dens}, and thus the inequality $\mathrm{WEC}_3$ is always satisfied in this region. 

The remaining inequalities, in regions where $A(x)>0$, are
\begin{equation}
    \begin{split}
        &NEC_{2}
\Longleftrightarrow -\frac{l_0^2 \left(6 l_0^5+13 l_0^3 x^2+x^4 (7 l_0-9 M)\right)}{16 \pi  x^4
   \left(l_0^2+x^2\right)^{5/2}}\\
&\qquad\qquad\quad+\frac{l_0^2\left(3 l_0^2+2  x^2\right)}{8 \pi x^4\left(l_0^2 + x^2\right)}\geq 0,
    \end{split}
\end{equation}
\begin{equation}
\begin{split}
&SEC_3 \Longleftrightarrow -\frac{l_0^2 \left(6 l_0^5+11 l_0^3 x^2+x^4 (5 l_0-3 M)\right)}{8 \pi  x^4
   \left(l_0^2+x^2\right)^{5/2}}\\
&\qquad\qquad\quad+\frac{l_0^2\left(3 l_0^2+ x^2\right)}{4 \pi x^4\left(l_0^2 +  x^2\right)}                           \geq 0,
\end{split}
\end{equation}
\begin{equation}
         DEC_{1} \Longrightarrow \frac{l_0^2 \left(\frac{\left(l_0^2+x^2\right) \left(\sqrt{l_0^2+x^2}-l_0\right)}{x^2}+3 M\right)}{4 \pi 
   \left(l_0^2+x^2\right)^{5/2}}\geq 0,
\end{equation}
\begin{equation}
\begin{split}
&DEC_{2} \Longrightarrow \frac{6 l_0^7+13 l_0^5 x^2+l_0^2 x^4 (7 l_0+15 M)}{16 \pi  x^4 \left(l_0^2+x^2\right)^{5/2}}\\
&\qquad\qquad\quad-\frac{l_0^2\left(3 l_0^2+2  x^2\right)}{8 \pi x^4 \left(l_0^2 +  x^2\right)}\geq 0. 
\end{split}
\end{equation}

In regions where $A(x)<0$ we find
\begin{equation}
    \begin{split}
        &NEC_{2}
\Longleftrightarrow -\frac{l_0^2 \left(6 l_0^5+17 l_0^3 x^2+x^4 (11
   l_0+3 M)\right)}{16 \pi  x^4 \left(l_0^2+x^2\right)^{5/2}}\\
 & \qquad\qquad\quad+ \frac{l_0^2\left(3 l_0^2+4  x^2\right)}{8 \pi x^4\left(l_0^2 + x^2\right)}\geq 0,
    \end{split}
\end{equation}
\begin{equation}
      SEC_3 \Longleftrightarrow\frac{-6 l_0^7-15 l_0^5 x^2-9 l_0^2 x^4 (l_0+M)}{8 \pi  x^4
   \left(l_0^2+x^2\right)^{5/2}}+\frac{3 l_0^2}{4 \pi  x^4}\geq 0,
\end{equation}
\begin{equation}
    DEC_{1} \Longrightarrow \frac{l_0^2 \left(\frac{\left(l_0^2+x^2\right) \left(\sqrt{l_0^2+x^2}-l_0\right)}{x^2}+3 M\right)}{4 \pi 
   \left(l_0^2+x^2\right)^{5/2}}\geq 0,
\end{equation}

\begin{eqnarray}
&&DEC_{2} \Longrightarrow \frac{6 l_0^7+9 l_0^5 x^2+3 l_0^2 x^4 (l_0+M)}{16 \pi  x^4 \left(l_0^2+x^2\right)^{5/2}}\nonumber\\
&&\qquad\qquad\quad-\frac{3 l_0^4}{8
   \pi  l_0^2 x^4+8 \pi  r^6}\geq 0, \\
&&WEC_3 \Longleftrightarrow \frac{l_0^2 \left(\sqrt{l_0^2+x^2}-l_0\right)}{4 \pi  x^2 \left(l_0^2+x^2\right)^{3/2}}\geq 0.
\end{eqnarray}
It is possible to see that $WEC_3$ is satisfied within the event horizon once $\sqrt{l_0^2+x^2}>l_0$. By the same argument, it can be seen that $DEC_1$ is always satisfied. With some care, it is possible to verify that the remaining energy conditions are satisfied in certain regions using the expressions above. However, to make this information clearer, we will graphically analyze the behavior of combinations of the stress-energy tensor components.

\begin{figure}[htb!]
    \centering
     \includegraphics[width=1\linewidth]{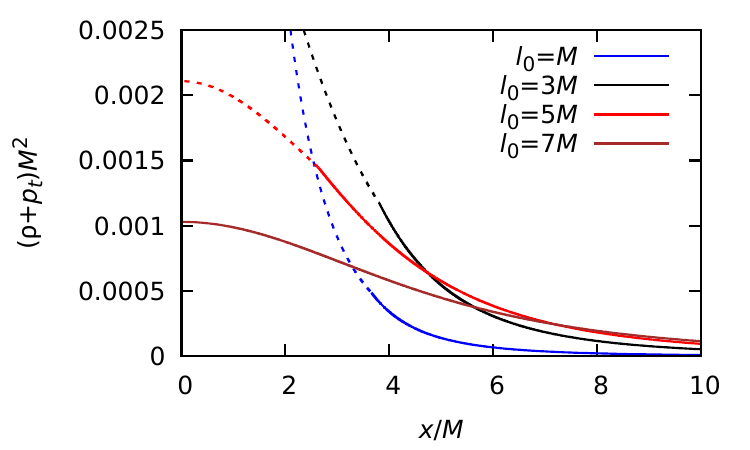} 
     \includegraphics[width=1\linewidth]{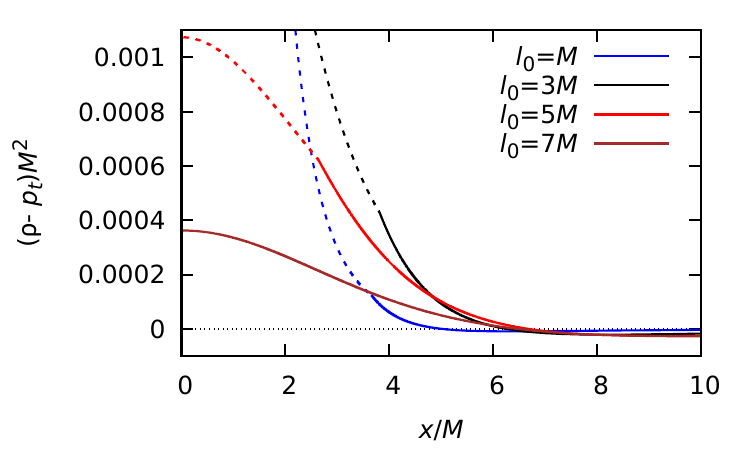}
     \includegraphics[width=1\linewidth]{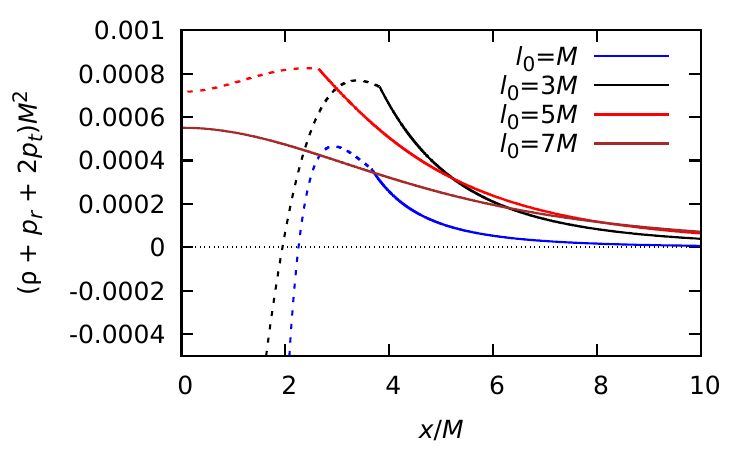}
    \caption{Combinations of the stress-energy tensor components as functions of the radial coordinate for different values of $l_0$. The solid part of each curve represents regions where $A(x) > 0$, while the dashed part corresponds to regions where $A(x) < 0$.}
    \label{fig:energy}
\end{figure}

 In Fig. \ref{fig:energy}, it can be seen that the inequality associated with $SEC_3$ is violated in the inner regions for some values of $l_0$, while for larger values of $l_0$, it appears to be always satisfied. The inequality associated with $DEC_2$ is violated at larger distances but is satisfied inside the horizon. The remaining inequalities are always satisfied. Therefore, the source of our solution is less exotic than other BB models found in the literature.

\section{Conclusion}\label{S:conclusion}
In this work, we investigated static and spherically symmetric BB spacetimes sourced by an effective energy density inspired by T-duality, which naturally introduces a minimal length scale acting as an ultraviolet regulator. Within the framework of classical general relativity, this effective matter content allowed us to construct regular geometries that smoothly interpolate between black holes and traversable wormholes, providing a concrete realization of the BB scenario.

To obtain the spacetime, we imposed the energy density arising from string-inspired corrections associated with T-duality and also prescribed a wormhole area. Using Einstein's equations together with the conservation of the stress-energy tensor, we derived the metric components and the stress-energy tensor that generate this spacetime. The matter content supporting the solution is described by an anisotropic fluid. The spacetime features a wormhole throat at $x=0$ and two event horizons, one in the region $x>0$ and the other in $x<0$. For $l_{0}<6M$, the solution corresponds to a regular black hole with two event horizons. When $l_{0}=6M$, the wormhole throat and the horizons coincide, yielding a black throat and thus a one-way traversable wormhole. For $l_{0}>6M$, no event horizon exists and the spacetime describes a two-way traversable wormhole. The causal structure of this spacetime is very similar to that of the Simpson--Visser geometry, as illustrated in Figs.~\ref{fig:diagramTW}, \ref{fig:oneway}, and \ref{fig:regular}. By analyzing the Kretschmann scalar, we verified the regularity of the spacetime and confirmed the absence of curvature singularities.

We also studied the geodesics in this spacetime. For massive particles, depending on the values of $\ell$ and $l_{0}$, it is possible to have one stable and one unstable circular orbit outside the event horizon. In the wormhole case, there is an additional stable orbit located at the wormhole throat, together with the symmetry of the orbits under the transformation $x\to -x$. We identified the presence of an innermost stable circular orbit (ISCO) and found that, initially, the ISCO radius increases as $l_{0}$ increases, reaching a maximum value $x_{\text{ISCO}}^{\max}\approx18.29M$ at $l_{0}\approx13.75M$. Beyond this point, the ISCO radius decreases and eventually ceases to exist for $l_{0}\approx31.05M$. For massless particles, we found that in the regular black hole regime there exists only a single circular photon orbit, which is unstable. In the wormhole case, however, the structure is richer: there are two unstable circular photon orbits, one located in the region $x>0$ and the other in $x<0$, together with a stable circular orbit at the throat, $x=0$. The unstable orbit in the region $x>0$ exists up to $l_{0}=12M$. Consequently, there is an interval $6M<l_{0}<12M$ in which no event horizons are present and all three photon orbits become accessible. For $l_{0}>12M$, only a single unstable circular orbit remains, located at $x=0$.

Using EHT observational data, it is possible to impose constraints on the parameters of our model. From the analysis of massless geodesics, we computed the shadow radius of the compact object described by our solution and compared it with the EHT measurements through the mass–radius relation. An important point emerges from this comparison: if one considers the ratio $r_{S}/M$, our model does not fit the EHT constraints. However, the appropriate quantity to use is the ADM mass, which corresponds to the mass measured by an observer at infinity. When we instead analyze the ratio $r_{S}/M_{\mathrm{ADM}}$, we find that our model is consistent with the EHT data at the $2\sigma$ level for $l_{0}\lesssim1.15\,M_{\mathrm{ADM}}$. In fact, when expressed in terms of the ADM mass, our solution reproduces the Schwarzschild case remarkably well.

Moreover, we have simulated the optical appearance for this class of spacetimes by considering a couple of cases for the free parameters that describe a regular black hole ($l_0/M=1$) and a wormhole ($l_0/M=7$), respectively. To do so, a Standard Unbound distribution is considered for the luminosity of the accretion disk. Although the scale and asymmetry of the intensity profile are fixed, different positions for the luminosity peak are considered to simulate the observed intensity. In order to compare this spacetime with some others, the Schwarzschild spacetime is also included as well as a particular case where the spacetime given in \cite{Lobo:2020ffi} provides a wormhole solution. The choice of the former and the latter is not causal, but both have some links to the BB found in this paper. Then, the optical appearance arises with special features in comparison to  Schwarzschild spacetime, particularly the shape and the width of the light ring structure of the images. Although the comparison among the wormholes shows that the case \cite{Lobo:2020ffi} has a much more complex structure due to the presence of a larger number of light rings, which just displays a direct consequence of the presence of several circular orbits for photons \cite{Guerrero:2022qkh}, whereas the wormhole featured by \eqref{Ax} contains a unique photon sphere (on each side of the throat), leading to a simpler structure of the light rings.

We also studied the thermodynamics associated with this solution in the regime where an event horizon is present. We found that the Hawking temperature increases as the horizon radius decreases, closely reproducing the Schwarzschild behavior for sufficiently large black holes. However, the temperature reaches a maximum value at a critical radius, where the heat capacity diverges, signaling a second-order phase transition, Figs. \ref{fig:T} and \ref{fig:C}. Beyond this point, the heat capacity becomes positive, indicating the emergence of a thermodynamically stable branch. In the small-horizon limit, the temperature decreases and approaches zero as $x_H\to 0$, suggesting that evaporation slows down indefinitely. Moreover, the Hernandez-Misner-Sharp mass does not vanish in this limit, but instead approaches a finite residual value $M_0=l_0/2$. Therefore, rather than complete disappearance, the evaporation process is expected to end in a cold remnant configuration, which can be interpreted as a regular compact object without an event horizon.

Finally, we analyzed the energy conditions associated with the model. For this purpose, we identified the components of the stress-energy tensor corresponding to an anisotropic fluid. The expressions depend on whether we are considering regions inside or outside possible horizons, i.e., $A>0$ or $A<0$. By examining the inequalities associated with the null energy condition, we found that $NEC_{1}$ is violated whenever $\Sigma>0$ and $\Sigma''>0$, which is precisely the case in our model. Therefore, the null energy condition is always violated, since at least one of its inequalities fails everywhere. The inequalities associated with $WEC_{3}$ and $DEC_{1}$ are always satisfied, while the remaining inequalities depend on the choice of parameters. Overall, our model is able to satisfy more inequalities than other BB scenarios, making it, to some extent, less exotic.

Our BB model opens up a wide range of new possibilities, since we now have a solution that does not rely on NED nor on a phantom scalar field to be sustained. In future work, we intend to investigate perturbations of this spacetime through the analysis of quasinormal modes and the time evolution of these perturbations, linking them to the light ring structure formed in the optical appearance of these objects \cite{Duran-Cabaces:2025sly}. We also plan to explore how rotation may influence the properties and physical implications of this model.

\section*{Acknowledgments}
The authors would like to thank Conselho Nacional de Desenvolvimento Cient\'{i}fico e Tecnol\'ogico (CNPq) and Funda\c c\~ao Cearense de Apoio ao Desenvolvimento Cient\'ifico e Tecnol\'ogico (FUNCAP) for the financial support. The work is also supported by the Spanish project PID2024-157196NB-I00 funded by MICIU/AEI/10.13039/501100011033; and the financial support by the Department of Education, Junta de Castilla y Le\'{o}n and FEDER Funds, 
Ref.~CLU-2023-1-05.

\bibliographystyle{apsrev4-1}
\bibliography{ref.bib}
\end{document}